\documentclass[letterpaper,10pt]{article}  
\usepackage{osameet2}
\usepackage{ae}	

\usepackage{graphicx}  
\usepackage{amsmath}                
\usepackage{amsfonts}   
\usepackage{bm}	
\usepackage{bbm}	
\usepackage{color}    
\usepackage{datetime}
\usepackage{xspace}
\usepackage[colorlinks=true,bookmarks=false,citecolor=blue,urlcolor=blue]{hyperref}
\usepackage{relsize}
\usepackage{textpos}
\usepackage{booktabs}

\newcommand{\degsym}{\ensuremath{^\circ}}    
\newcommand{\hairsp}{\hspace{1pt}}
\newcommand{\ie}{\textit{i.\hairsp{}e.}\xspace}
\newcommand{\eg}{\textit{e.\hairsp{}g.}\xspace}
\newcommand{\foreign}[1]{\emph{#1}}    
\newcommand{\code}[1]{\texttt{#1}}    
\newcommand{\acr}[1]{\textsc{\smaller{#1}}}    
\newcommand{\psdi}[1]{\ensuremath{#1^{+}}}        
\newcommand{\vctr}[1]{\ensuremath{\boldsymbol{#1}}}        
\newcommand{\adj}[1]{\ensuremath{#1^{t}}}        

\begin{document}
\begin{textblock}{1.}[.5,0.](0.5,-.1)
 \noindent Preprint of paper \href{http://www.osa.org/en-us/meetings/optics_and_photonics_congresses/classical_optics/international_optical_design_conference_\%28iodc\%29/invited_speakers/}{IM3A.1} of the \textit{International Optical Design Conference (\acr{IODC})}, Fairmont Orchid, Kohala Coast, Hawaii, \acr{USA}  (2014).\hfill\tiny rev \pdfdate\normalsize\\\rule{\textwidth}{.5pt}\\[.25in]
\end{textblock}

\title{Ray-tracing for coordinate knowledge\\ in the \acr{JWST}\\ Integrated Science Instrument Module}

\author{Derek Sabatke, Joseph Sullivan}
\address{Ball Aerospace \& Technologies Corp., \acr{PO} Box 1062, Boulder \acr{CO} 80306}
\email{corresponding author's e-mail: dsabatke@ball.com}

\author{Scott Rohrbach, David Kubalak}
\address{\acr{NASA} Goddard Space Flight Center, Greenbelt Rd., Greenbelt, \acr{MD} 20771}

\begin{abstract}
Optical alignment and testing of the Integrated Science Instrument Module of the James Webb Space Telescope is underway. We describe the Optical Telescope Element Simulator used to feed the science instruments with point images of precisely known location and chief ray pointing, at appropriate wavelengths and flux levels, in vacuum and at operating temperature. The simulator's capabilities include a number of devices for in situ monitoring of source flux, wavefront error, pupil illumination, image position and chief ray angle.  Taken together, these functions become a fascinating example of how the first order properties and constructs of an optical design (coordinate systems, image surface and pupil location) acquire measurable meaning in a real system.  We illustrate these functions with experimental data, and describe the ray tracing system used to provide both pointing control during operation and analysis support subsequently. Prescription management takes the form of optimization and fitting.  
Our core tools employ a matrix/vector ray tracing model which proves broadly useful in optical engineering problems.  We spell out its mathematical basis, and illustrate its use in ray tracing plane mirror systems relevant to optical metrology such as a pentaprism and  corner cube.
\end{abstract}

\ocis{080.0080, 0080.273, 080.2468}

\section{overview}
\label{sect:overview}

\begin{figure}
\begin{center}
\includegraphics[width=5.58in]{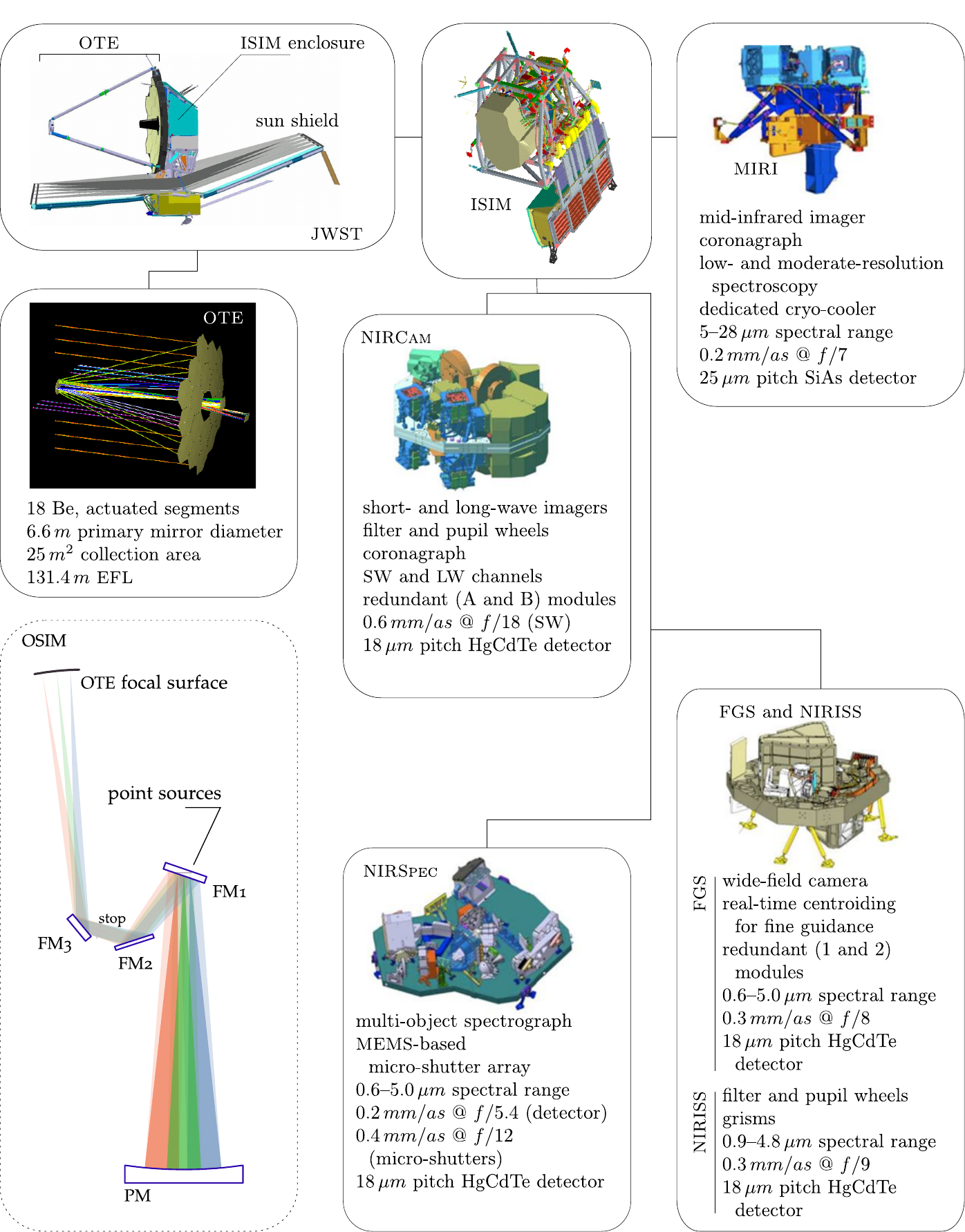}
\caption{\label{fig:JWSTOpticalOverview}Taxonomy of major \acr{JWST} optical subsystems.  \acr{OSIM} is ground support equipment which substitutes for the \acr{OTE} during cryo-vacuum testing of \acr{ISIM}. In the listed plate scales, $as$ stands for arc seconds.}
\end{center}
\end{figure}

The James Webb Space Telescope (\acr{JWST})\cite{Lightsey2012} is an infrared observatory under construction by an international collaboration including \acr{NASA}, the European Space Agency (\acr{ESA}), and the Canadian Space Agency (\acr{CSA}).
Featuring a $6.6\,m$ collection aperture diameter and spectral sensitivity in the $0.6$--$27\,\mu m$ wavelength range, \acr{JWST} will enable studies of the early universe and extra-solar planetary systems.
The major \acr{JWST} optical subsystems include the optical telescope element (\acr{OTE}) and the integrated science instrument module (\acr{ISIM}), as illustrated in Figure~\ref{fig:JWSTOpticalOverview}.
        
The \acr{OTE} is \acr{JWST}'s objective telescope.
It features a segmented primary mirror assembly which is stowed for launch and deployed and aligned in flight.
The 18 hexagonal primary mirror segments have Be substrates with protected Au coatings and feature tip/tilt, translation and radius-of-curvature actuation for fine alignments. 
The image surface is divided into allocations for each of the 5 science instruments\cite{Davila2004}: \acr{NIRCam}, \acr{NIRSpec}, \acr{MIRI}, \acr{FGS} and \acr{NIRISS}.
These are mounted in \acr{ISIM}'s composite structure, clustering the instruments' pick-off mirrors and field stops near the \acr{OTE} focal surface. 
    
The \acr{OTE} and \acr{ISIM} are currently following independent integration and test paths, to be merged for combined alignment and testing once complete.
The \acr{OTE}'s Be primary mirror segments have been fabricated, figured, coated and verified at temperature.  Integration of the primary mirror segment assembly backplane is on-going.
    
All 5 science instruments have been completed, individually tested, delivered and integrated into \acr{ISIM}. A series of cryo-vacuum (\acr{CV}) tests \cite{Kimble2012} are under way at \acr{NASA}'s Goddard Space Flight Center to align the instruments in \acr{ISIM}, verify focus, pupil and image alignment, demonstrate co-boresight stability, calibrate residual wavefront errors and verify a number of other requirements at the module level. 
The first of these tests (designated \acr{CV1rr} for \emph{risk reduction}) featured only \acr{MIRI}, \acr{FGS} and \acr{NIRISS}, and was completed in late 2013.
Two additional tests (\acr{CV2} and \acr{CV3}) are planned, with \acr{CV2} commencing in the summer of 2014.

In order to conduct optical tests of \acr{ISIM} without the \acr{OTE} present, a \acr{CV}-compatible optical relay has been constructed.
Known as the Optical Telescope Element Simulator (\acr{OSIM}), it serves as the master tool for the alignment and calibration of \acr{ISIM}.
In this paper we focus our attention on facets of \acr{OSIM}'s use which fall outside the usual scope of optical metrology discussions, but may be of particular interest to the optical design community. 
These include ray-tracing topics related to how we estimate \acr{OSIM}'s optical prescription with emphasis on image and pupil location, how we make use of \foreign{in situ} alignment tools to update that prescription to maintain absolute coordinate knowledge, and how we interface that knowledge to other optical design and analysis tools.  
To put this material in context, we begin with an overview of \acr{OSIM} in Section~\ref{Sect:OSIM}.

In Section~\ref{Sect:RayTracingMechanics} we develop a ray tracing framework based on linear algebra.
While the underpinnings of this approach are hardly novel, it proves not only useful for compact and robust computer code in \acr{OSIM} operations, but valuable for broader optical engineering insights as well.  
Section~\ref{Sect:OSIMSoftware} gives an overview of the ray tracing infrastructure developed for \acr{OSIM}, and Section~\ref{Sect:PointCal} describes how the underlying \acr{OSIM} prescription is calibrated and maintained. Finally Section~\ref{Sect:Results} gives a selection of test results illustrating the application of these tools in the tests to date.

\section{Optical Telescope Element Simulator\label{Sect:OSIM}}
\subsection{Optical Bench Module}
\acr{OSIM} is a finite-conjugate, optical relay which enables simultaneous, \acr{CV} stimulus of all \acr{ISIM} science instruments with accurate exit pupil and image coordinates\cite{Davila2008, Sullivan2010}.
Its basic optical design is similar to a Schmidt telescope with the corrector plate omitted.  It has a single powered optic --- the $1.8\,m$ diameter primary mirror (\acr{PM}) --- and a stop located close to the center of curvature (Figure~\ref{fig:JWSTOpticalOverview}).
Three fold mirrors (\acr{FM1}--\acr{FM3}) follow the \acr{PM} for beam delivery.
Sources take the form of optical fibers or back-lit pinholes on a source plate close behind \acr{FM1}.
Their light is injected through small holes in \acr{FM1}, which act as obscurations in the beam.
An additional obscuration is caused by a center hole in the \acr{PM} through which the Alignment Diagnostics Module (\acr{ADM}) views.

The relative simplicity of the optical design aids in alignment and prescription diagnostics.  
A tolerable $40\,nm$ \acr{RMS} of uncorrected spherical aberration remains in the design on-axis, which couples into astigmatism and coma for off-axis field points.
These as-built aberrations are calibrated by phase retrieval and interferometry, and account is taken of them when characterizing the science instruments.
The supporting bench structure is Al, which undergoes considerable shrinkage between room and \acr{CV} temperatures (shifting focus by nearly $9\,cm$).
The impacts of this cost-saving measure have been mitigated by cryogenic calibration and use of separate warm and cold prescriptions.

With the exception of \acr{FM3}, the mirrors are made from \foreign{Zerodur}\texttrademark and have protected Al coatings.
\acr{FM3} has a fused silica substrate, a dichroic coating on the front and an anti-reflection coating on the back to transmit a visible-wavelength return beam to support pupil alignments.
Care was required in the coating designs to prevent polarization effects from affecting the laser ranging system of the \acr{ADM}\cite{Sabatke2009}.

Several \acr{OSIM} components have mechanical actuators to facilitate steering across instruments' fields of view, resulting in 8 actuated degrees of freedom (\acr{DoF}).
In the nominal alignment, the source plate has a source for each science instrument module (plus one on-axis to mark the master chief ray).
This enables stimulation of multiple instruments simultaneously, to simulate guiding and investigate co-boresight stability at the so-called \emph{fixed field points}.
The Source Plate Mechanism (\acr{SPM}) is actuated in 3 \acr{DoF} of translation in order to sweep focus.
It sees additional use for small dithers in field, and to select which of several fibers or pinholes in the source bundles are moved to nominal.

Broadly speaking, simultaneous stimulation of multiple science instruments is only possible at the fixed field points.
(Other scenarios can be conceived, but necessitate compromises in focus or pupil position for at least one of the provided output beams.)
For individual instrument tests, it's necessary to be able to provide a steerable image, which is accomplished via 2 \acr{DoF} tip/tilt actuation of \acr{FM3}.
Such actuation impacts focus, which is compensated with \acr{SPM} motion.
\pagebreak
It also impacts exit pupil position since the stop is not quite coincident with \acr{FM3}.
This is compensated with 3 \acr{DoF} translation actuation of the stop via the pupil translation mechanism (\acr{PTM}).

\subsection{Alignment Diagnostics Module}
\begin{figure}
\begin{center}
\includegraphics[width=5.53in]{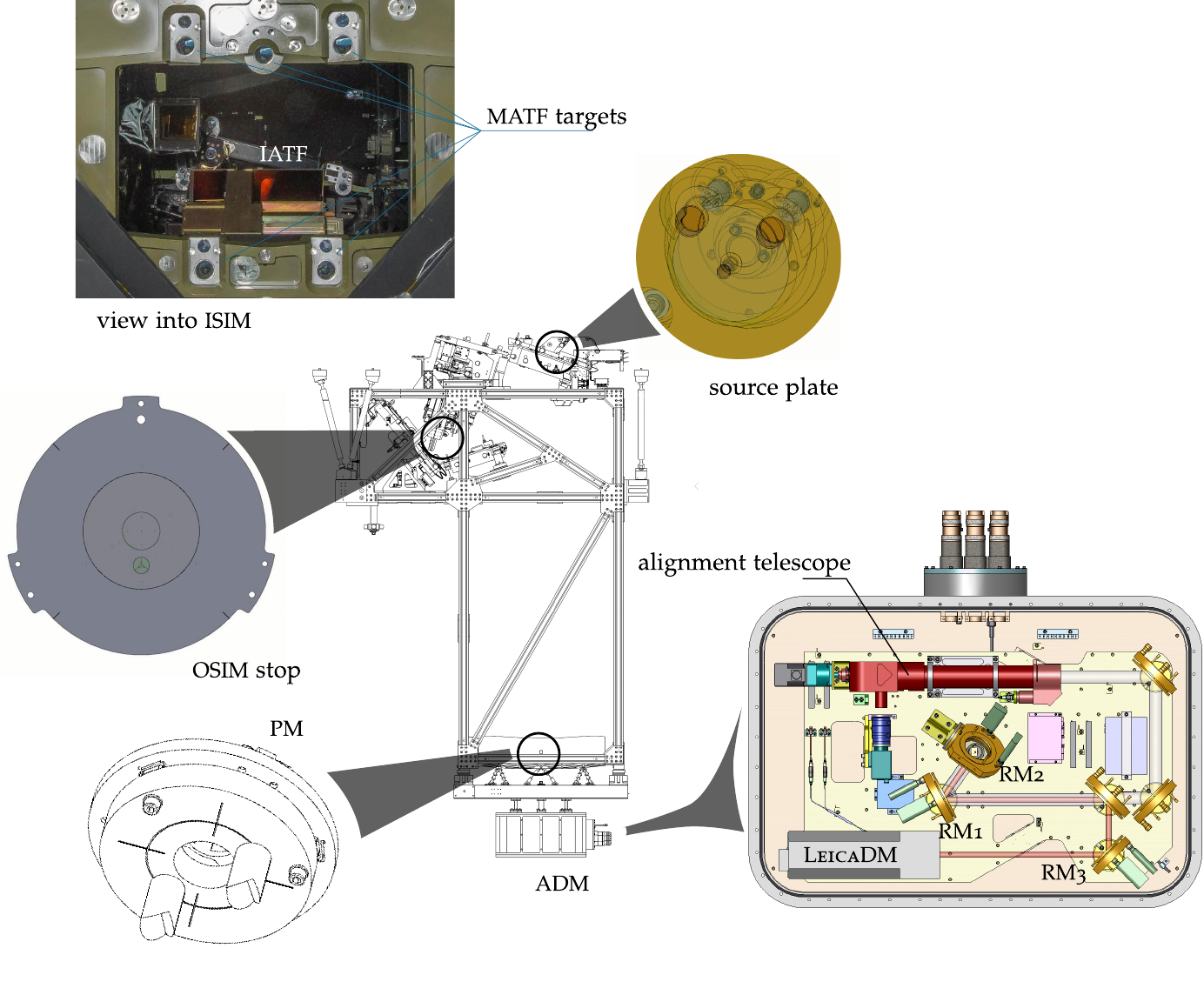}
\caption{\label{fig:ADM}The alignment diagnostics module and a selection of its targets.}
\end{center}
\end{figure}

Located beneath the \acr{OSIM} \acr{PM} is the \acr{ADM}, an optical metrology system featuring an alignment telescope and ranging laser.
The \acr{ADM} instruments operate at room temperature and atmospheric pressure from within a pressure vessel, looking into the \acr{OSIM} optical system through a vacuum window and a hole in the center of the \acr{PM} (Figure~\ref{fig:ADM}).
The alignment telescope can be used either as an autocollimator (sighting the surface normal of a distant flat mirror) or a spotting telescope (sighting distant point targets such as back-lit pinholes or target fiducials).  
A number of such targets are provided, including internal targets for nulling the \acr{ADM} instruments, targets on the \acr{OSIM} bench, and metrology targets on two test fixtures downstream of \acr{OSIM}.  

The illuminated reticle for the autocollimator features an extended pattern of grid lines and a binary designation system to distinguish between grid cells\cite{Leviton2003}.
This provides a mechanism for reading autocollimation errors quantitatively if the system needs to be used off-axis (without nulling to the target's normal vector).
This is not our typical mode of operation.  Where possible, we use \acr{FM3} to null the return and use the motion recorded by \acr{FM3}'s angular encoders as our reading.
Nonetheless, the ability to read autocollimator returns quantitatively is useful for finding a null.
Having an extended illumination source for alignment telescope operations also proves quite useful.
We show example images in Section~\ref{Sect:RayTracingMechanics} in which the reticle provides a convenient test image for image orientation and parity discussions. 

The ranging laser system (a \foreign{Leica Absolute Distance Meter}\texttrademark) is nominally boresighted to the alignment telescope by means of a dichroic beam combiner within the \acr{ADM}.
Ranging measurements are facilitated by return cubes (usually solid glass retroreflectors) on the target assemblies.
The two systems' beams are independently steerable by means of remotely controlled mirrors within the pressure vessel.
In most cases ranging measurements are in fact made with the laser steered off from nominal to avoid the center hole in \acr{FM1} for the master chief ray source.
Generally this obscuration does not interfere with alignment telescope measurements.
The exception is spotting the target in the \acr{OSIM} stop, which is close enough to \acr{FM1} to be obscured on axis.
That difficulty was worked around by using 4 symmetrically displaced pinholes for the pupil target rather than one on-axis pinhole.

\subsection{calibration standards}
The Master Alignment Test Fixture (\acr{MATF}) resides between \acr{OSIM} and \acr{ISIM}, and provides spotting, autocollimation and ranging targets just outside the \acr{ISIM} field-of-view allocations.
A limited set of additional targets is available on the \acr{ISIM} Alignment Test Fixture (\acr{IATF}), which mounts to \acr{ISIM} and is removed before flight.
The \acr{MATF} is our main set of references for determining updates to the \acr{OSIM} prescription using \acr{ADM} measurements.
Absolute coordinates for \acr{MATF} and \acr{IATF} targets are known through a combination of laboratory metrology and \acr{CV} qualification in a photogrammetry-equipped system (prior to testing with \acr{ISIM})\cite{Nowak2010,Hadjimichael2010}. 
In addition to alignment targets, the \acr{IATF} holds a point diffraction interferometer (\acr{PDI}) and radiometrically calibrated detectors for monitoring \acr{OSIM}'s sources.
The \acr{PDI} provides the capability of measuring wavefront error as well as obtaining pupil intensity images.

During \acr{OSIM} qualification, \acr{ISIM} was replaced by the Beam Image Analyzer (\acr{BIA}).
It featured a metrologized phase retrieval camera on translation actuators which were calibrated to absolute coordinates.
The \acr{BIA} could therefore traverse the region near the \acr{OTE} focal surface and provide absolute image position and chief ray angle measurements for an initial \acr{CV} tuning of \acr{OSIM}'s prescription.
The \acr{BIA} also had a \acr{PDI} and flux monitors.

\subsection{Pupil Imaging Module}
Each science instrument has a pupil alignment reticle (\acr{PAR}) located approximately in a pupil.
A science instrument \acr{PAR} is a flat mirror with registration features imprinted.
To mitigate any stray light risk they are designed to fit within the central obscuration of the \acr{JWST} pupil.
\acr{OSIM} has similar transmissive shadow masks in its pupil wheels.

\begin{figure}
\begin{center}
\includegraphics[width=3.82in]{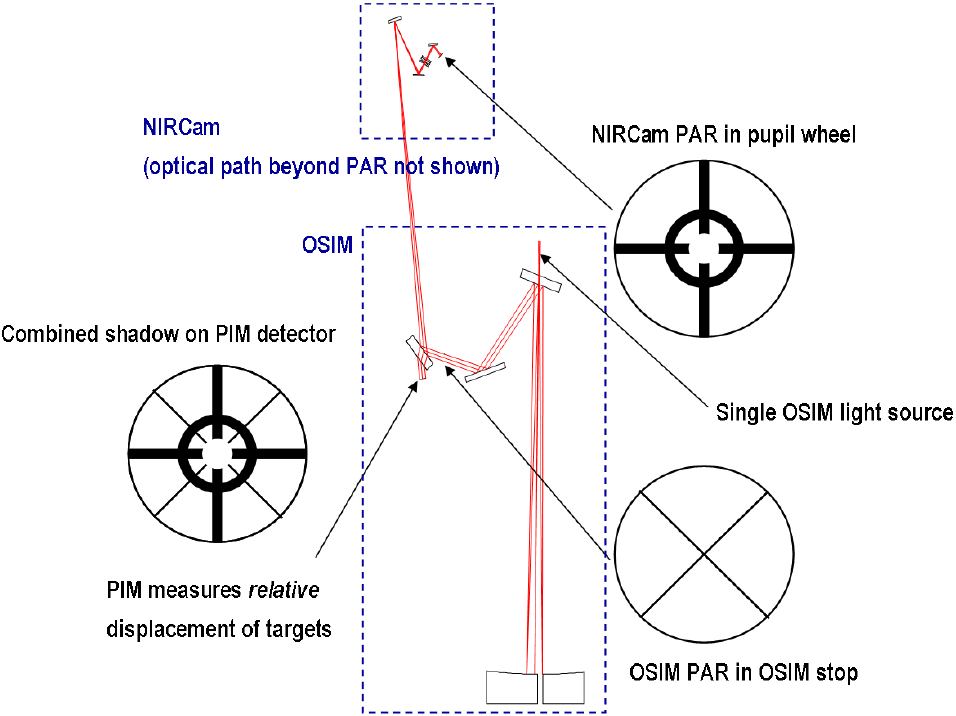}
\caption{\label{fig:PIMPAR}The \acr{OSIM} pupil imaging module records the reflected return from a fiducialized flat mirror near a science instrument pupil. }
\end{center}
\end{figure}

When \acr{OSIM} illuminates the science instrument with an appropriate field point, the instrument's \acr{PAR} generates a return beam on which both the \acr{OSIM} and science instrument reticles are imprinted (Figure~\ref{fig:PIMPAR}).
The intensity pattern of the return beam is recorded after transmitting through \acr{FM3} on its return path by a bare detector, designated the Pupil Imaging Module (\acr{PIM}).
Image analysis is employed to determine the offset between masks, which indicates any pupil misalignments.\looseness=1 

In addition, the \acr{NIRCam} short wave channel has a pupil imaging lens. 
When deployed, an image of the \acr{OSIM} pupil can be obtained directly on the \acr{NIRCam} shortwave detectors.

\section{ray tracing mechanics}
\label{Sect:RayTracingMechanics}
\subsection{operator formalism}
We find it useful to employ a vector-based ray tracing model.  In this formalism a ray is represented by a point in space (through which the ray is known to pass) and a unit vector giving the ray's direction.
Many of the ray-tracing and coordinate calculations required in operating and analyzing data from \acr{OSIM} are conveniently represented with vectors and vector operators (see \eg \cite{BarrettImageScience}).  
The use of coordinate algebra in ray tracing is hardly novel.  In fact, the topic is so well developed that it's often treated in a computer science context rather than an optical one\cite{Dorst2007}.
However this formulation proves useful in both analytic and numerical exercises, and warrants description.

We will represent a vector in boldface, adding a caret for a unit vector (\eg $\vctr{\hat{v}}$). 
Generally we will use a transpose to denote a dot product ($\vctr{u}\cdot\vctr{v}\equiv\adj{\vctr{u}}\vctr{v}$) since the notation lends itself to an operator framework.
In a particular coordinate system, a vector $\vctr{v}$ may be represented by a 3-element column vector, a functional $\adj{\vctr{u}}$ takes the form of a 3-element row vector, and a general linear operator takes the form of a $3\times3$ matrix. (Note that the operator $\vctr{v}\adj{\vctr{v}}$, which will be encountered shortly, may be computed as a matrix multiplication of a $3\times1$ column vector and a $1\times3$ row vector as the notation suggests.  However in numerical analysis environments such as \foreign{Mathematica}\texttrademark\ and \foreign{Python}\cite{Python} it may be more conveniently implemented using outer product routines.) 

Wherever possible we avoid explicit designation of a coordinate system or expansion into rows and columns.
This allows us to carry the analytic math forward in an elegant fashion.
Even when coding for numerical evaluation, avoiding dependence on specific choices of coordinate system simplifies the implementation and notation, reduces the risk of coding errors, and makes future maintenance easier.

As a first operator example, consider the projection of a vector $\vctr{v}$ onto a unit vector $\hat{\vctr{d}}$.  
We haven't specified a context for this operation yet, but it is one that is ubiquitous in physics (\eg in resolving forces into components or any type of vector into a coordinate system).  
The length of the projection is given by the scalar product $\adj{\hat{\vctr{d}}}\vctr{v}$.  The resultant direction is $\hat{\vctr{d}}$, so the projected vector is 
\begin{equation}
\vctr{v}_\parallel=(\adj{\hat{\vctr{d}}}\vctr{v})\hat{\vctr{d}} = (\hat{\vctr{d}}\adj{\hat{\vctr{d}}})\vctr{v}
\label{LinearProjectionEqn}
\end{equation}
In the second form of the equation we've taken advantage of associativity to group the expression into the form of a projection operator $\hat{\vctr{d}}\adj{\hat{\vctr{d}}}$ acting on the vector $\vctr{v}$.
This notation is much more intuitive than it may at first seem.
Reading this operator notation right-to-left, it immediately reminds us of the action of the operator --- to perform a dot product with $\hat{\vctr{d}}$ and then multiply the resulting scalar into the vector $\hat{\vctr{d}}$ to obtain a result.

To obtain the component of the vector \emph{perpendicular} to $\hat{\vctr{d}}$ we simply subtract the projection:
\begin{equation}
\vctr{v}_\perp=\vctr{v}-\hat{\vctr{d}}\adj{\hat{\vctr{d}}}\vctr{v} = (\mathbbm{1}-\hat{\vctr{d}}\adj{\hat{\vctr{d}}})\vctr{v}
\end{equation}
Here $\mathbbm{1}$ is the identity operator, and the operator $\mathbbm{1}-\hat{\vctr{d}}\adj{\hat{\vctr{d}}}$ projects onto the subspace perpendicular to $\hat{\vctr{d}}$. Such an operation may be familiar to students of quantum physics or signal processing\cite{BarrettImageScience,CohenTannoudji1977,Mallat1999}.
The ability to specify \emph{everything but} a dimension given by a unit vector is the source of much of the convenience and elegance of our operator/vector approach to ray tracing.

A ray (which for our purposes is no different than a line) is specified by giving a point $\vctr{r}$ in space (though which the ray is known to pass) and a direction vector $\hat{\vctr{d}}$ (which points parallel to the ray).
Consider the problem of projecting a point $\vctr{\rho}$ onto the ray $(\vctr{r}, \hat{\vctr{d}})$.
Position vectors are unlike other types (such as direction, velocity, force, \foreign{etc.}) in their reference to a coordinate origin.
The difficulty vanishes for displacement vectors, so we can project the displacement from $\vctr{r}$ to $\vctr{\rho}$ as
\begin{equation}
\vctr{\rho}_\mathrm{proj}-\vctr{r}=(\hat{\vctr{d}}\adj{\hat{\vctr{d}}})(\vctr{\rho}-\vctr{r})
\end{equation}

Having to subtract the reference point $\vctr{r}$ on both sides of this equation may at first seem like a liability.  
Wouldn't it be nicer if the projection operator could be applied to $\vctr{\rho}$ alone (as in Equation~\ref{LinearProjectionEqn})?
In a computer science context\cite{Dorst2007}, a $4\times4$ transformation operator is introduced, which incorporates translations and changes of origin. 
However, here we opt for a different tack, since the issue is essential.
In coordinate metrology, mathematical choice of origin should never matter to a final result.  
Physical meaning lies in the distances between points in the system under test --- not in metrology equipment or the reference format of the data.
Moreover, implicit origins are a communications hazard, so it's good practice to avoid them anyway. 
Therefore, having to subtract reference points can be viewed as explicitly specifying an origin, and therefore as an advantage to this approach rather than a disadvantage.

Table~\ref{tbl:VectorRayTracing} summarizes definitions and operators for a selection of basic ray tracing components.
It's not necessary to spell out each in detail here, but we will touch on highlights and describe a few applications.

\begin{table}[h]
\caption{basic constructs and operations for vector ray tracing}
\label{tbl:VectorRayTracing}
  \begin{center}
    \begin{tabular}{rl}
    \toprule
    \midrule
    displacement from $\vctr{r}$ to $\vctr{\rho}$ & $\vctr{\rho}-\vctr{r}$ \\
    normalization (of any vector $\vctr{v}$) & $\vctr{\hat{v}}=\vctr{v}/\sqrt{\adj{\vctr{v}}\vctr{v}}$ \\
    included angle $\theta$ between unit vectors & $\cos\theta = \adj{\hat{\vctr{u}}}\hat{\vctr{v}}$ \\
    \cmidrule(lr){1-2}
    projection operator for direction $\hat{\vctr{d}}$ & $\hat{\vctr{d}} \adj{\hat{\vctr{d}}}$ \\
    projection operator for subspace $\perp$ to $\hat{\vctr{d}}$ & $\mathbbm{1}-\hat{\vctr{d}}\adj{\hat{\vctr{d}}}$ \\
    rotation operator (angle $\alpha$, right handed about axis $\hat{\vctr{a}}$) & $\hat{\vctr{a}}\adj{\hat{\vctr{a}}}+\cos\alpha(\mathbbm{1}-\hat{\vctr{a}}\adj{\hat{\vctr{a}}})+\sin\alpha(\hat{\vctr{a}}\times\mathbbm{1})$ \\
    \cmidrule(lr){1-2}
    definition of a ray (\ie line) & point $\vctr{r}$, direction $\vctr{\hat{d}}$ \\
    parameterization of a ray (signed distance $s$) & $\vctr{\rho}(s)=\vctr{r} + s\,\hat{\vctr{d}}$ \\
    projection of point $\vctr{\rho}$ onto ray $(\vctr{r},\vctr{\hat{d}})$ & $\vctr{r}+\vctr{\hat{d}}\adj{\vctr{\hat{d}}}(\vctr{\rho}-\vctr{r})$ \\
    $\perp$ from a ray $(\vctr{r},\vctr{\hat{d}})$ to a point $\vctr{\rho}$ & $(\mathbbm{1}-\hat{\vctr{d}}\adj{\hat{\vctr{d}}})(\vctr{\rho}-\vctr{r})$ \\
    \cmidrule(lr){1-2}
    definition of a plane & point $\vctr{p}$, normal $\hat{\vctr{n}}$\\
    intersection of a ray $\vctr{r}+s\,\vctr{\hat{d}}$ and plane $(\vctr{p},\vctr{\hat{n}})$ & $s = {(\vctr{p}-\vctr{r})\cdot\vctr{\hat{n}}}/{\vctr{\hat{d}}\cdot\vctr{\hat{n}}}$ \\  
    reflection of a ray  $(\vctr{r},\vctr{\hat{d}})$ in a plane  $(\vctr{p},\vctr{\hat{n}})$ & ${\hat{\vctr{d}}}'=(\mathbbm{1}-2\ \hat{\vctr{n}}\adj{\hat{\vctr{n}}})\hat{\vctr{d}}$\\
    & $(\vctr{r}'-\vctr{p})=(\mathbbm{1}-2\ \hat{\vctr{n}}\adj{\hat{\vctr{n}}})(\vctr{r}-\vctr{p})$\\
    \cmidrule(lr){1-2}
    definition of a sphere & center $\vctr{c}$, radius $R$\\
    intersection(s) of a ray $\vctr{r}+s\,\vctr{\hat{d}}$ with a sphere $(\vctr{c},R)$ & $s = -(\vctr{r}-\vctr{c})\cdot\vctr{\hat{d}}$\\
    &$\hphantom{s=\ \ }\pm\sqrt{R^2-(\vctr{r}-\vctr{c})\cdot(\vctr{r}-\vctr{c})+[(\vctr{r}-\vctr{c})\cdot\vctr{\hat{d}}]^2}$ \\
    \bottomrule
    \end{tabular}
  \end{center}
\end{table}

\subsection{rotations and coordinate axis construction}
Consider the rotation operator 
\begin{equation}
\label{RotOpCrossEqn}
R(\alpha,\hat{\vctr{a}})=\hat{\vctr{a}}\adj{\hat{\vctr{a}}}+\cos\alpha(\mathbbm{1}-\hat{\vctr{a}}\adj{\hat{\vctr{a}}})+\sin\alpha(\hat{\vctr{a}}\times\mathbbm{1})
\end{equation}
which rotates the vector on which it operates by angle $\alpha$ in a right-handed sense about an axis $\hat{\vctr{a}}$.
The operator can be readily interpreted term-by-term.
The first term indicates that the component of its operand which is parallel to the axis is passed through unchanged.
The next term attenuates the orthogonal component by a factor of $\cos\alpha$, and the final term constructs a mutually orthogonal basis vector and erects a component scaled by $\sin\alpha$.
In the third term, $\hat{\vctr{a}}\times\mathbbm{1}$ denotes the operator which takes the cross-product with $\hat{\vctr{a}}$ on the left.
In terms of explicit matrices, if in a particular coordinate system $\hat{\vctr{a}}$ has the representation $\hat{\vctr{a}}| = (a_x, a_y, a_z)$, then in the same system
\begin{equation}
\hat{\vctr{a}}\times\mathbbm{1}|=
\begin{pmatrix} 
0 & a_z & -a_y \\
-a_z & 0 & a_x \\
a_y & -a_x & 0 \\
\end{pmatrix}
\end{equation}

An alternate form of this operator can be obtained once we assign names to axes orthogonal to $\hat{\vctr{a}}$.  Calling them $\hat{\vctr{b}}$ and $\hat{\vctr{c}}$ (where we take the system $\hat{\vctr{a}}$, $\hat{\vctr{b}}$, $\hat{\vctr{c}}$ to be right-handed) we have
\begin{equation}
\label{RotOpSysEqn}
R(\alpha,\hat{\vctr{a}})=\hat{\vctr{a}}\adj{\hat{\vctr{a}}}+\cos\alpha\ \hat{\vctr{b}}\adj{\hat{\vctr{b}}}-\sin\alpha\ \hat{\vctr{b}}\adj{\hat{\vctr{c}}}+\sin\alpha\ \hat{\vctr{c}}\adj{\hat{\vctr{b}}} + \cos\alpha\ \hat{\vctr{c}}\adj{\hat{\vctr{c}}}
\end{equation}
This is an element-by-element expression for the operator in the $\hat{\vctr{a}}$, $\hat{\vctr{b}}$, $\hat{\vctr{c}}$ system, whose matrix representation in that system takes on the familiar form for a rotation matrix
\begin{equation}
R(\alpha,\hat{\vctr{a}})|_\textrm{a,b,c}=
\begin{pmatrix} 
1 & 0 & 0 \\
0 & \cos\alpha & -\sin\alpha \\
0 & \sin\alpha & \cos\alpha \\
\end{pmatrix}
\end{equation}

The operator form in Equation~\ref{RotOpSysEqn} has the disadvantage that it requires construction of a coordinate system, but the advantage (over Equation~\ref{RotOpCrossEqn}) of avoiding a cross-product.
When construction of a coordinate system is required, we often find it convenient to encapsulate the task with a subroutine (named \code{GenSys} in our implementations) to construct the system using a Gram-Schmidt process\cite{BarrettImageScience}.
This nicely handles a common situation in which we want one coordinate axis exactly parallel to a known vector (\eg a chief ray direction), another axis to lie in the plane of that vector and a second selection (\eg to lie in the plane of our chief ray and the default $y$ axis), and the final vector to preserve orthogonality and handedness.
A practical \code{GenSys} implementation should have the ability to fill in dimensions if fewer than 3 vectors are given, to re-order dimensions in case we don't want the preserved vector to be the $x$ axis, and automatically apply right-handedness after re-ordering.
Canned \acr{QR} decomposition\cite{Golub1996} routines can be useful here, since normalization and over-specification of input vectors can be handled automatically. 
Then by simply concatenating all three coordinate system axes in every construction we avoid a cross-product's indeterminacy in the case when parallel vectors are given.

It's worth emphasizing the difference between rotation \emph{operators} (which have matrix representations) and coordinate transformation matrices (which may embody coordinate systems defined in terms of rotations).  
A rotation operator transforms the vector on which it operates, returning a different vector (in the same coordinate system, when working with matrix representations).
A coordinate transform matrix returns a representation of the same vector in a new system.
When a rotation is used to define a new coordinate system by rotating the axes of an existing one, the matrices representing the rotation and the coordinate transformation are different but related.  
Confusing them often results in changing the sign of the intended rotation --- an error which may not be apparent at the conceptual level but is potentially devastating to numerical results.

Defining a coordinate system in terms of Euler angles could be done as follows.  Suppose we have an initial system $\hat{\vctr{x}}$, $\hat{\vctr{y}}$, $\hat{\vctr{z}}$, and want to define a new system $\hat{\vctr{x}}'$, $\hat{\vctr{y}}'$, $\hat{\vctr{z}}'$ in terms of $xyz$ Euler angles.  In our operator notation this can be written
\begin{equation}
\label{EulerRotationEquation}
(\hat{\vctr{x}}'\ \hat{\vctr{y}}'\ \hat{\vctr{z}}') = R(\theta_z, \hat{\vctr{z}})R(\theta_y,\hat{\vctr{y}})R(\theta_x,\hat{\vctr{x}}) (\hat{\vctr{x}}\ \hat{\vctr{y}}\ \hat{\vctr{z}})
\end{equation}
Here the notation is meant to indicate that the composite rotation operator is applied to each of the initial coordinate axes $\hat{\vctr{x}}$, $\hat{\vctr{y}}$, $\hat{\vctr{z}}$ in turn.  
In a matrix implementation, this can be viewed as a matrix-matrix multiplication.
If $\hat{\vctr{x}}$, $\hat{\vctr{y}}$, $\hat{\vctr{z}}$ is already the working system, $(\hat{\vctr{x}}\ \hat{\vctr{y}}\ \hat{\vctr{z}})|$ is the identity matrix, and the columns of the composite rotation operator are directly the new system's coordinate axes.
It's important to note that this is not by any means the only set of Euler angles that sees use.
In modeling with \acr{Code V} in particular the $zyx$ angles are called for (with additional sign flips for some of the angles to arrive at \code{ADE}, \code{BDE}, \code{CDE} decenters).
These are defined as
\begin{equation}
\label{ZYXEulerRotationEquation}
(\hat{\vctr{x}}'\ \hat{\vctr{y}}'\ \hat{\vctr{z}}') = R(\phi_x,\hat{\vctr{x}}) R(\phi_y,\hat{\vctr{y}})R(\phi_z, \hat{\vctr{z}})(\hat{\vctr{x}}\ \hat{\vctr{y}}\ \hat{\vctr{z}})
\end{equation}
Because of the complexities of dealing with Euler angles, we strongly recommend communicating coordinate systems with explicit lists of vector axes, and avoiding external references to Euler angles entirely.
A subroutine to convert Euler angles to axis vectors is an important component of practical ray tracing calculations.
Delineating the basis of such a routine is outside our scope here, but we note that Equations~\ref{EulerRotationEquation} and \ref{ZYXEulerRotationEquation} are key.
For example, by substituting into Equation~\ref{ZYXEulerRotationEquation} it can be shown that the $zyx$ Euler angles given by
\enlargethispage{2\baselineskip}
\begin{equation}
\label{ZYXEulerIdentityEquation}
\begin{array}{rcr}\phi_x'&=&-(\pi-\phi_x)\\
\phi_y'&=&\pi-\phi_y\hphantom{)}\\
\phi_z'&=&-(\pi-\phi_z)
\end{array}
\end{equation}produce the same coordinate system.

That is, more than one set of Euler rotations may produce a given coordinate system, and branch cuts in a conversion routine may produce mathematically-correct but unconventional results. 
In such cases, the transformation in Equation~\ref{ZYXEulerIdentityEquation} may be used to correct the Euler angles.

\subsection{plane mirrors}
A plane may be specified by a point $\vctr{p}$ (through which the plane is known to pass) and a normal vector $\hat{\vctr{n}}$.
Note that this specification is somewhat similar to that of a line, but here the vector is orthogonal to the shape of interest --- not along it.
Projection into the plane and construction of the orthogonal residual employ the projection operators constructed form the normal.

An operator describing reflection by a plane (mirror) is easily constructed by noting that the law of reflection stipulates that the component of the input vector parallel to the plane is unchanged, and the component normal to it has its sign reversed.  
We use the operator $\hat{\vctr{n}}\adj{\hat{\vctr{n}}}$ to identify the normal component.
Subtracting that operator from the identity operator (as in projections) nulls the component, so we subtract with a factor of 2 to reverse it.
Thus the reflection operator is given by
\begin{equation}
M(\hat{\vctr{n}})=\mathbbm{1}-2\,\hat{\vctr{n}}\adj{\hat{\vctr{n}}}
\end{equation}
To reflect a ray $(\vctr{r},\hat{\vctr{d}})$ we can apply this operator directly to the direction vector $\hat{\vctr{d}}$ since it does not reference a coordinate origin.
To obtain the reflected point, we apply the operator to the displacement from $\vctr{r}$ to any point in the plane (as in Table~\ref{tbl:VectorRayTracing}).

It's worth noting that the reflection operator for a point and ray are the same.
This is necessary (or perhaps obvious) for imaging to be correctly represented.
That is, any ray can be represented as a set of points, so a ray must obey the reflection properties of points.
Similarly, a point is the intersection of two rays, so a point must obey the reflection properties of rays.

These results are derived specifically for plane mirrors.  The reflection operator for a ray may be extended to a curved mirror surface by using the local normal and vertex point at the ray's intersection with the surface.
 
\subsection{systems of plane mirrors}

When describing a system of plane mirrors, the reflection operators can be multiplied together to obtain a net operator.  
(Propagating the ray points is not quite so elegant, requiring daisy chaining the offsets.)
For analysis of autocollimator data in which only ray bearing and mirror normals matter, this is a very convenient result.
Measurements made through systems of mirrors (including fold flats and pentaprisms) can be \emph{exactly} reduced once the beam delivery system's normal vectors are known.
Emphasis is placed on exactness here, since there's no need to rely on 2-dimensional approximations and assumptions.

Note that a reflection operator is it's own inverse
\begin{equation}
(\mathbbm{1}-2\hat{\vctr{n}}\adj{\hat{\vctr{n}}})^2=\mathbbm{1}
\end{equation}
and the reflection operator for a mirror is the same regardless of the sign of the normal
\begin{equation}
 \mathbbm{1}-2(-\hat{\vctr{n}})(-\adj{\hat{\vctr{n}})} = \mathbbm{1}-2\hat{\vctr{n}}\adj{\hat{\vctr{n}}}
\end{equation}
With these results it's trivial to show that a system of 2 parallel plane mirrors displace a beam without changing its direction (regardless of the direction of the incoming beam).

\subsection{example: a 2 mirror system}

Consider a two mirror system, the first with normal $\hat{\vctr{n}}_1=\hat{\vctr{n}}$ and the second with normal $\hat{\vctr{n}}_2=\cos\theta \hat{\vctr{n}}+\sin\theta \hat{\vctr{m}}$ where $\theta$ is the angle between the mirrors and we have resolved $\hat{\vctr{n}}_2$ into components in the plane of the normals.
Specifically, we use our \code{GenSys} routine on inputs $\hat{\vctr{n}}_1$, $\hat{\vctr{n}}_2$ to construct coordinate vectors $\hat{\vctr{n}}$, $\hat{\vctr{m}}$, $\hat{\vctr{a}}$, where $\hat{\vctr{n}}$ and $\hat{\vctr{m}}$ span the plane of the normals and $\hat{\vctr{a}}$ is perpendicular to them.
The product
\begin{equation}
 M_\textrm{net}=(\mathbbm{1}-2\, \hat{\vctr{n}}_2\adj{\hat{\vctr{n}}_2})(\mathbbm{1}-2\, \hat{\vctr{n}}_1\adj{\hat{\vctr{n}}_1})
\end{equation} gives the net system operator.
Plugging in the expansions for $\hat{\vctr{n}}_2$ and applying the completeness identity 
\begin{equation}
 \mathbbm{1}=\hat{\vctr{n}}\adj{\hat{\vctr{n}}}+ \hat{\vctr{m}}\adj{\hat{\vctr{m}}}+\hat{\vctr{a}}\adj{\hat{\vctr{a}}} \mbox{\rule{0.25in}{0pt}(in 3 dimensions)}
\end{equation} and trigonometric simplifications, we find
\begin{equation}
 M_\textrm{net}=\hat{\vctr{a}}\adj{\hat{\vctr{a}}}+\cos(2\theta)\hat{\vctr{n}}\adj{\hat{\vctr{n}}}-\sin(2\theta)\hat{\vctr{n}}\adj{\hat{\vctr{m}}}+\sin(2\theta)\hat{\vctr{m}}\adj{\hat{\vctr{n}}}+\cos(2\theta)\hat{\vctr{m}}\adj{\hat{\vctr{m}}}
\end{equation} Comparing this result to Equation~\ref{RotOpSysEqn} we see that the action of the system is a rotation through an angle $2\theta$ about an axis parallel to the (perhaps virtual) intersection of the two planes.
This is a well-known result\cite{Smith2007}  in two dimensions.
The operator version incorporates the third dimension readily, and indicates that the out-of-plane component of the incoming ray direction vector propagates unperturbed.

This is a potentially significant result in optical metrology setups using pentaprisms (for which $2\theta=90\degsym$).
In those cases, the measurement is insensitive to misalignments of the pentaprism by rotations about $\hat{\vctr{a}}$, but susceptibility to alignment remains for the other two degrees of freedom.
This result shows that it's possible to tolerance those modes of misalignments (or even avoid their impact entirely by applying an exact ray trace in the data analysis) using little more than a matrix multiplication.
                    
\subsection{example: application to a corner cube}
Analyzing the action of a (hollow) corner cube turns out to be nearly trivial with an appropriate choice of coordinate system.
For reflection by a single mirror, if we take the mirror normal to be the $z$ axis of our working coordinate system, the reflection operator is represented by
\begin{equation}
 (\mathbbm{1}-2\,\hat{\vctr{n}}\adj{\hat{\vctr{n}}})|_{\hat{\vctr{z}}=\hat{\vctr{n}}}=
 \begin{pmatrix}
 1 & 0 & 0 \\
 0 & 1 & 0 \\
 0 & 0 & -1
 \end{pmatrix}
\end{equation} 

Proceeding similarly for the second reflection (choosing the normal as the $y$ axis) and third reflection (the $x$ axis) we have for the net reflection operator for a corner cube,
\begin{equation}
 M_\mathrm{cc}=
 \begin{pmatrix}
 -1 & 0 & 0 \\
 0 & 1 & 0 \\
 0 & 0 & 1
 \end{pmatrix}
 \begin{pmatrix}
 1 & 0 & 0 \\
 0 & -1 & 0 \\
 0 & 0 & 1
 \end{pmatrix}
 \begin{pmatrix}
 1 & 0 & 0 \\
 0 & 1 & 0 \\
 0 & 0 & -1
 \end{pmatrix}
 = -\mathbbm{1}
\end{equation}

There are 6 different sequences in which a ray can traverse the three reflective surfaces of a corner cube.
Strictly speaking, the matrix ordering in this product applies to only one of them.
Since we are dealing with diagonal matrices however, they commute and clearly the net operator is the same for all 6 paths.
Moreover, if we choose our local coordinate origin at the vertex (the only point all three surfaces have in common) we find the same inversion holds for points.
Thus, \emph{a corner cube inverts rays and points about the vertex.}
The outgoing ray is antiparallel to the incoming ray.  
The output point is diametrically opposed to the incoming point.

\subsection{imaging properties of a corner cube} 
\begin{figure}
\begin{center}
\includegraphics[width=4.66in]{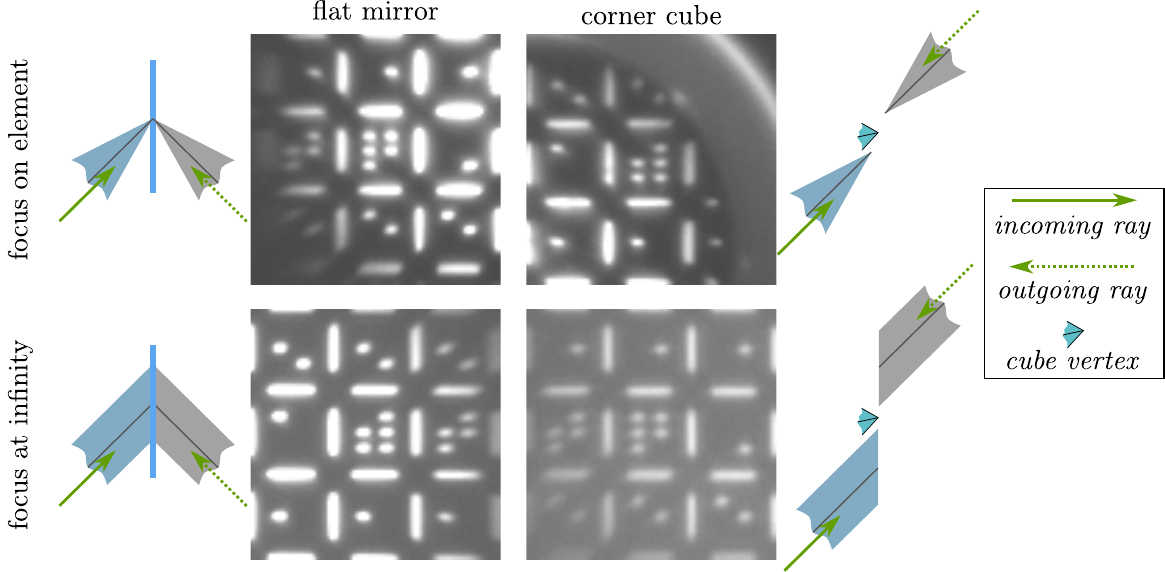}
\caption{\label{fig:CCImaging}Autocollimator return images from flats and corner cubes when focused on the element and at infinity.  Finite focus on a corner cube is analogous to infinity focus on a flat.}
\end{center}
\end{figure}

While these basic results for rays traversing a corner cube are well known, it may be worth pointing out a corner cube's imaging properties.
If we use an image-forming system (such as an autocollimator focused near the vertex) to input an optical image, the output image appears rotated $180\degsym$ about the vertex and shifted in focus symmetrically (Figure~\ref{fig:CCImaging}).

All 4 image cases in Figure~\ref{fig:CCImaging} exhibit plate scale of unit magnitude. 
Moving either along a row or column rotates the return image by $180\degsym$.  
There's a strong analogy along diagonals. 
It's easily seen from the figure that image orientation is preserved. 
The upper left (focus on a flat) and lower right (collimation on a corner cube) show \emph{cat's eye} insensitivity to target perturbations (respectively tilts of the flat or lateral translations of the corner cube).

On the other diagonal, a collimated return from a flat behaves much like a return from focus with a corner cube.  
In particular, both show a factor of two in sensitivity.  
Just as tilting an autocollimator flat by an angle $\theta$ introduces a return ray deviation of $2\theta$, offsetting a corner cube by a lateral distance $\delta$ translates the return image by $2\delta$.
Thus a corner cube is a potentially advantageous target for finite-conjugate spotting and triangulation measurements. 
In addition to a factor of two sensitivity versus a cross-hair or pinhole, it is easy to use with only the illumination source built into an autocollimator and is compatible with ranging devices such as the \acr{LeicaDM}.

\subsection{best-fit intersection of a ray bundle}
\begin{figure}
\begin{center}
\includegraphics[width=2.91in]{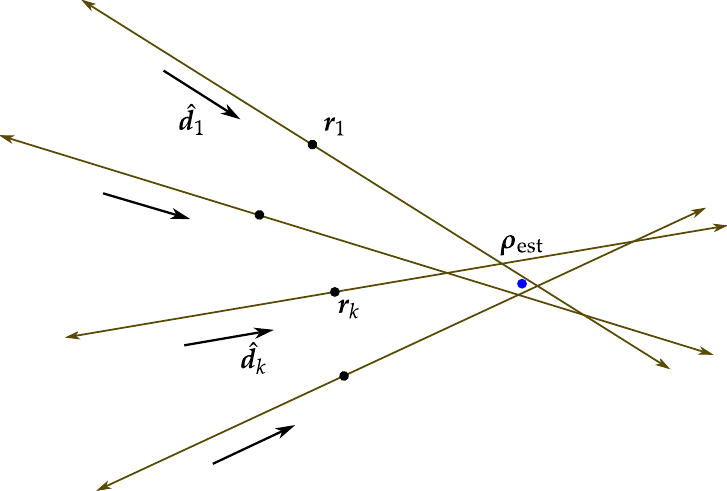}
\caption{\label{fig:RayBundleFit}Least-squares fitting to estimate the point of intersection of a ray bundle is carried out by means of the pseudoinverse.}
\end{center}
\end{figure}

Consider the task of finding the best-fit point of intersection of a bundle of rays (as in Figure~\ref{fig:RayBundleFit}).
This question is pertinent to a variety of tasks: finding the intersection of two rays, finding the point of closest approach of two skew rays, finding an estimate of focus from an entire bundle of rays exiting an optical system, and even a 2D image processing example to follow.
It's also included as an example of the use of the Moore-Penrose pseudoinverse \cite{BarrettImageScience} in classical optical engineering --- a topic worthy of considerable treatment in itself.

With each ray given by a point and a direction as $(\vctr{r}_k,\hat{\vctr{d}}_k)$ and a candidate intersection point $\vctr{\rho}$, the residual for each ray (\ie the perpendicular from each ray to $\vctr{\rho}$) is ideally zero:
\begin{equation}
 (\mathbbm{1}-\hat{\vctr{d}}_k\adj{\hat{\vctr{d}}_k})(\vctr{\rho}-\vctr{r}_k)=0 \mbox{\rule{12mm}{0mm}(ideally)}
\end{equation} so that
\begin{equation}
 (\mathbbm{1}-\hat{\vctr{d}}_k\adj{\hat{\vctr{d}}_k})\vctr{\rho}= (\mathbbm{1}-\hat{\vctr{d}}_k\adj{\hat{\vctr{d}}_k})\vctr{r}_k
\end{equation} In matrix representation in $N$ dimensions, this takes the form $H_k \vctr{\rho} = \vctr{g}_k$ of an $N\times{N}$ matrix of data multiplying the $N$-dimensional vector $\vctr{\rho}$ to be found, set equal to an $N\times1$ vector of data.
We have one such equation for each line in the bundle (indexed $k=1\textrm{\ldots}M$), and can stack them to give a combined statement $H \vctr{\rho}=\vctr{g}$.
Here $H$ is an $M N \times N$ data matrix and $g$ is an $M N \times 1$.
Thus we estimate the best-fit point of intersection as\clearpage
\begin{equation}
 \vctr{\rho}_\textrm{est}=\psdi{H}\vctr{g}
 \label{RayBundleFitEqn}
\end{equation} where $\psdi{H}$ is the Moore-Penrose pseudoinverse\cite{BarrettImageScience} of $H$.

\begin{figure}[b]
\begin{center}
\includegraphics[width=3.79in]{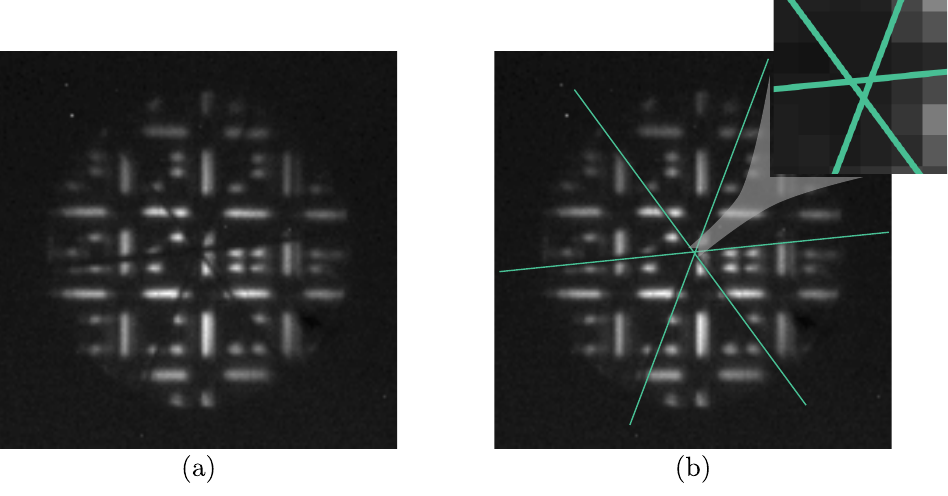}
\caption{\label{fig:CCVertexFit}An \acr{ADM} image focused at the vertex of a corner cube (a), with dihedral lines overlaid (b) to find the vertex location using Equation~\ref{RayBundleFitEqn}.}
\end{center}
\end{figure}

In addition to ray tracing and metrology applications in 3\acr{D}, we found this result useful in analyzing spotting telescope images (as in Figure~\ref{fig:CCVertexFit}(a)) from the \acr{ADM} when sighting at the apex of a return cube.
Illumination in this scenario comes from the grid and bit patterns of the autocollimator reticle, so that the three vertex lines can be made out at various places along their paths, but not necessarily at their intersection.
Using a drawing tool in an image analysis program, it is fairly easy to lay out lines along the vertices and export their coordinates (Figure~\ref{fig:CCVertexFit}(b)).
The analysis indicated in Equation~\ref{RayBundleFitEqn} (applied in $N=2$ dimensions) then gives a coordinate in pixel units for the vertex in the image.

\section{software tools for \acr{OSIM} and \acr{ISIM}\label{Sect:OSIMSoftware}}
\subsection{steering infrastructure}
We needed capability in the \acr{OSIM} control system to carry out ray-tracing calculations, in order to convert actuator encoder readings to and from image and pupil locations.
The control system runs in a tightly controlled environment, since it interacts with critical ground support equipment and flight hardware and must be extremely robust.
Very limited mathematical support was available from within the system itself, interfacing to external ray tracing engines would be difficult, and cross-platform compatibility was required.
To address theses needs, we designed a set of software routines to perform the required ray tracing calculations using the matrix/vector framework of Section~\ref{Sect:RayTracingMechanics} in the \foreign{Python} programming language and readily available extensions for mathematical analysis (such as \foreign{Numpy}\cite{Numpy} and \foreign{Scipy}\cite{Scipy}).

\begin{figure}
\begin{center}
\includegraphics[width=5.99in]{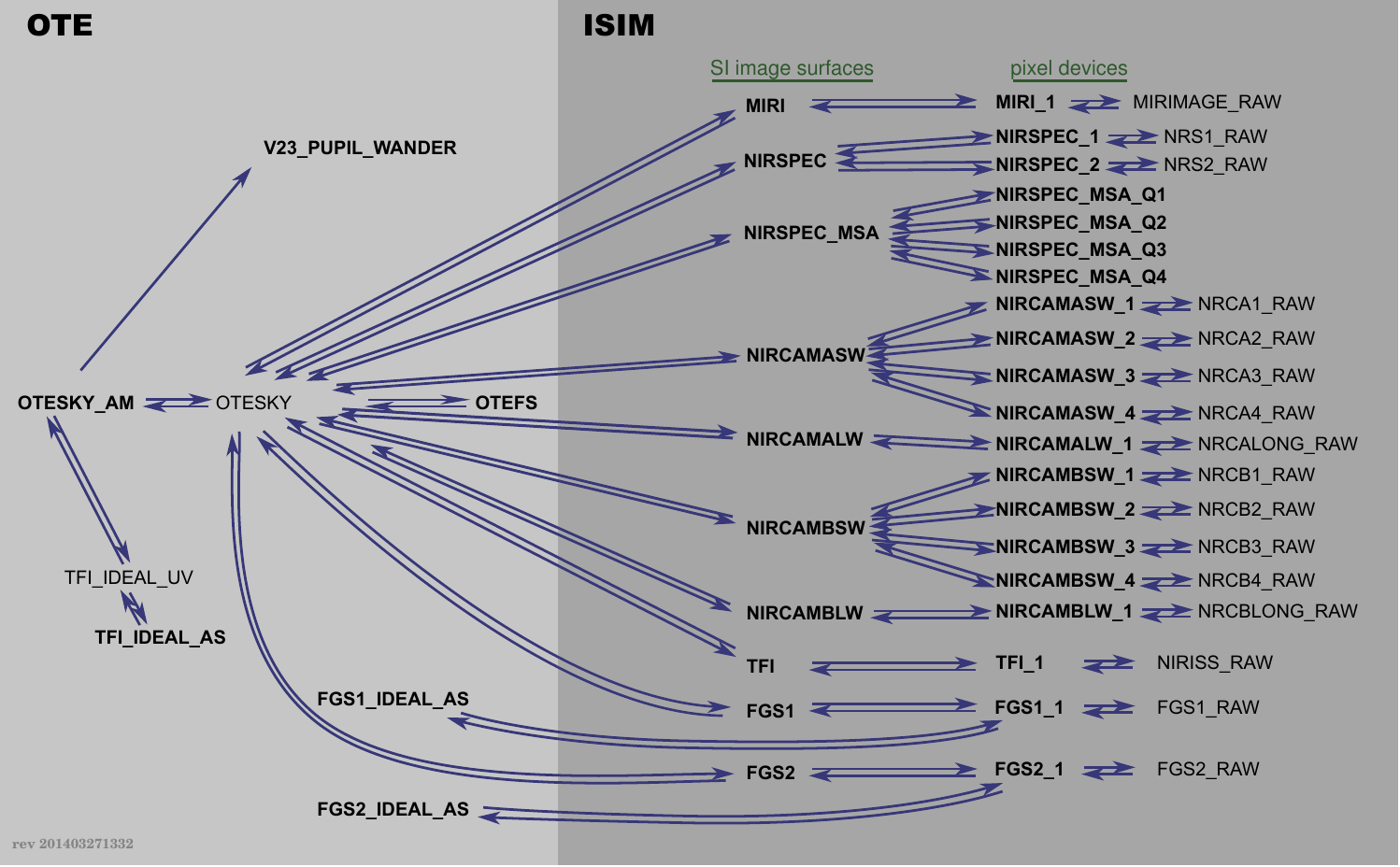}
\caption{\label{fig:CoordMapNetwork}\acr{OSIM} prescriptions contain a network of polynomial coordinate transformations for field points, to allow conversions among the numerous systems in which a field location may be specified.}
\end{center}
\end{figure}

To accomplish this, the steering routines need opto-mechanical data for \acr{OSIM} which includes surface data (such as positions and normals) and actuator calibration data (to convert from encoder readings to positions and orientation for the \acr{OSIM} mechanisms).
Ray tracing calculations generally begin or end at the \acr{OTE} focal surface.
However, for convenient management of the many tests planned for \acr{ISIM}, a host of different coordinate systems for specifying field point are supported.
These include pixel addresses in any of the 16 detector arrays in the various science instruments, sky coordinates, and control system coordinates for the fine guidance sensors.
A network of coordinate systems for field points is maintained along with the opto-mechanical data in an \acr{OSIM} prescription, which is passed to the steering code as a self-contained file.
Routing through this network is accomplished by means of the \foreign{NetworkX} graph theory package\cite{NetworkX}.
As implemented, the network has a star configuration, so there is only one route between any two systems (Figure~\ref{fig:CoordMapNetwork}), although loops are potentially useful (for example to support high-resolution transforms which are not defined over the entire range of a low-resolution counterpart).
This provides a modular framework, in which a stable code base is maintained for steering calculations but prescriptions can be revised as calibration data becomes available.

\begin{figure}
\begin{center}
\includegraphics[width=6in]{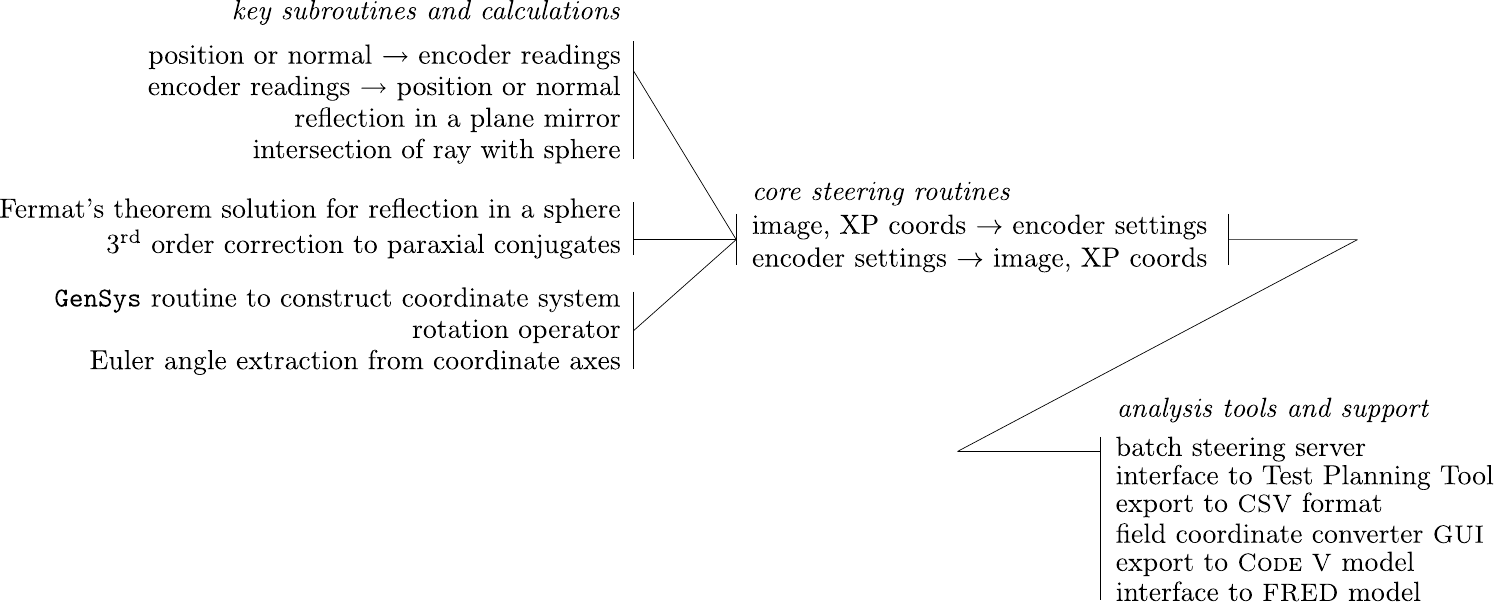}
\caption{\label{fig:SteerSoftware}Block-level components of the \acr{OSIM} steering software.}
\end{center}
\end{figure}

The core steering routines' interfaces are oriented specifically for operational control of \acr{OSIM}, and are intentionally very basic.
A number of other interfaces and wrappers have been developed for use by analysts and other users in test design and as-run data analysis.
These include routines to output data directly to \acr{Code V} and \acr{FRED} models using steering data in science image headers, embedding of the steering code in the \acr{ISIM} Optics Test Planning Tool, and a steering server which stays resident in system memory to speed up repeated steering calls during batch execution and analysis.
Figure~\ref{fig:SteerSoftware} summarizes this organization.

\section{calibrating \acr{OSIM} pointing\label{Sect:PointCal}}
With infrastructure in place for steering \acr{OSIM} using a prescription, the questions arise of how we obtain the prescription in the first place and maintain it during the life of the system.  
One of the main motivations for our approach to steering is precisely to facilitate these activities.
Initially the prescription was populated with design values.  
As data became available during integration and test --- first as \foreign{Romer}\texttrademark arm readings from mirror surfaces, later as theodolite and laser tracker data from the integrated \acr{OSIM}, finally as self-check and image location results from operational tests of \acr{OSIM} and the \acr{ADM}  --- the prescription was updated as necessary for consistency.
Moving forward through the test program, \acr{ADM} readings are used to check for signs of prescription changes, and apply prescription corrections if necessary.
The prescription becomes a container in which to accumulate knowledge of \acr{OSIM}'s alignment so that pointing is carried out with the best-available knowledge.

The variety of calibration data and cross-checks which could be used to test and if necessary update a prescription is immense.
For this reason we adopt a least-squares fitting model.
The prescription is used to predict expected outcomes for whatever measurements are available, predicted and actual outcomes are compared, and the prescription is adjusted as necessary to achieve agreement (to within expected uncertainties).

There are two main and recurring cases for which analysis workflows were developed.
These were calibration of \acr{OSIM} pointing with the \acr{BIA}, and prescription updates using the \acr{ADM} and \foreign{in situ} alignment targets in \acr{OSIM}, the \acr{MATF} and the \acr{IATF}.
The first scenario was limited to integration and qualification testing of \acr{OSIM}, since use of the \acr{BIA} precludes the presence of \acr{ISIM}.
The second scenario occurs routinely to maintain pointing accuracy while testing \acr{ISIM}.

For each scenario, we establish a least-squares fitting infrastructure.
This requires us to identify the type of measurement data available, provide analysis infrastructure to make corresponding predictions from a prescription, establish a way of quantifying the difference between measurement and prediction (\ie the residuals), establish a merit function to combine multiple residuals into a single value, establish a meaningful way to display residuals in order to advise changes to be made to the prescription, and establish an optimization loop which includes prescription perturbation and assessment steps in each iteration. 
The process is similar --- but not identical --- in both scenarios, and analogous to curve fitting tasks ubiquitous in science and engineering.
Perhaps the most important insight to be drawn from the analogy is the importance of inspecting residuals while plotted on their own scale.
Systematic behavior in the residuals is the most valuable clue available to the analyst in figuring out how the model needs to be adjusted, and these may be masked when plotting measurements and models together, rather than their difference.

\subsection{\acr{BIA} data and fitting}
\begin{figure}
\begin{center}
\includegraphics[width=2.61in]{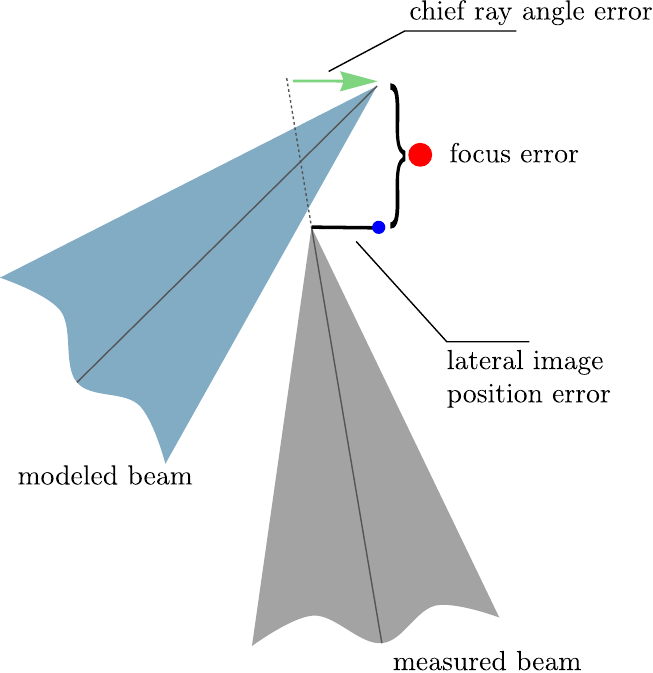}
\caption{\label{fig:RayResidualExplanation}An error in coordinate knowledge for an image-forming beam has 5 dimensions.  In order to display these conveniently on a \acr{2D} plot, they are broken into a \acr{2D} lateral image position error, a focus error and a \acr{2D} chief ray angle error using the symbols shown.  }
\end{center}
\end{figure}

For the case of \acr{BIA} calibrations, our measured data comprises image coordinates and  chief ray direction cosines for a number of selected field points.
Predictions of these values can be obtained from a prescription using the \acr{OSIM} steering code directly.
We found it most useful to present the residuals in a single 2\acr{D} plot of the \acr{ISIM} field of view, with graphical symbols (Figure~\ref{fig:RayResidualExplanation}) showing the magnitude and orientation of the error. 
Lateral pointing errors are shown with a black line, anchored at the nominal field location of the test case and the length indicating the error.  
Axial (\ie focus) errors are shown with a red square or circle (depending on the sign of the error), with the full width indicating the magnitude.
Chief ray errors are indicated by green arrows, their length showing the included angle between measurement and prediction, and their orientation showing the 2\acr{D} projection of the bearing error.

\begin{figure}
\begin{center}
\includegraphics[width=6.16in]{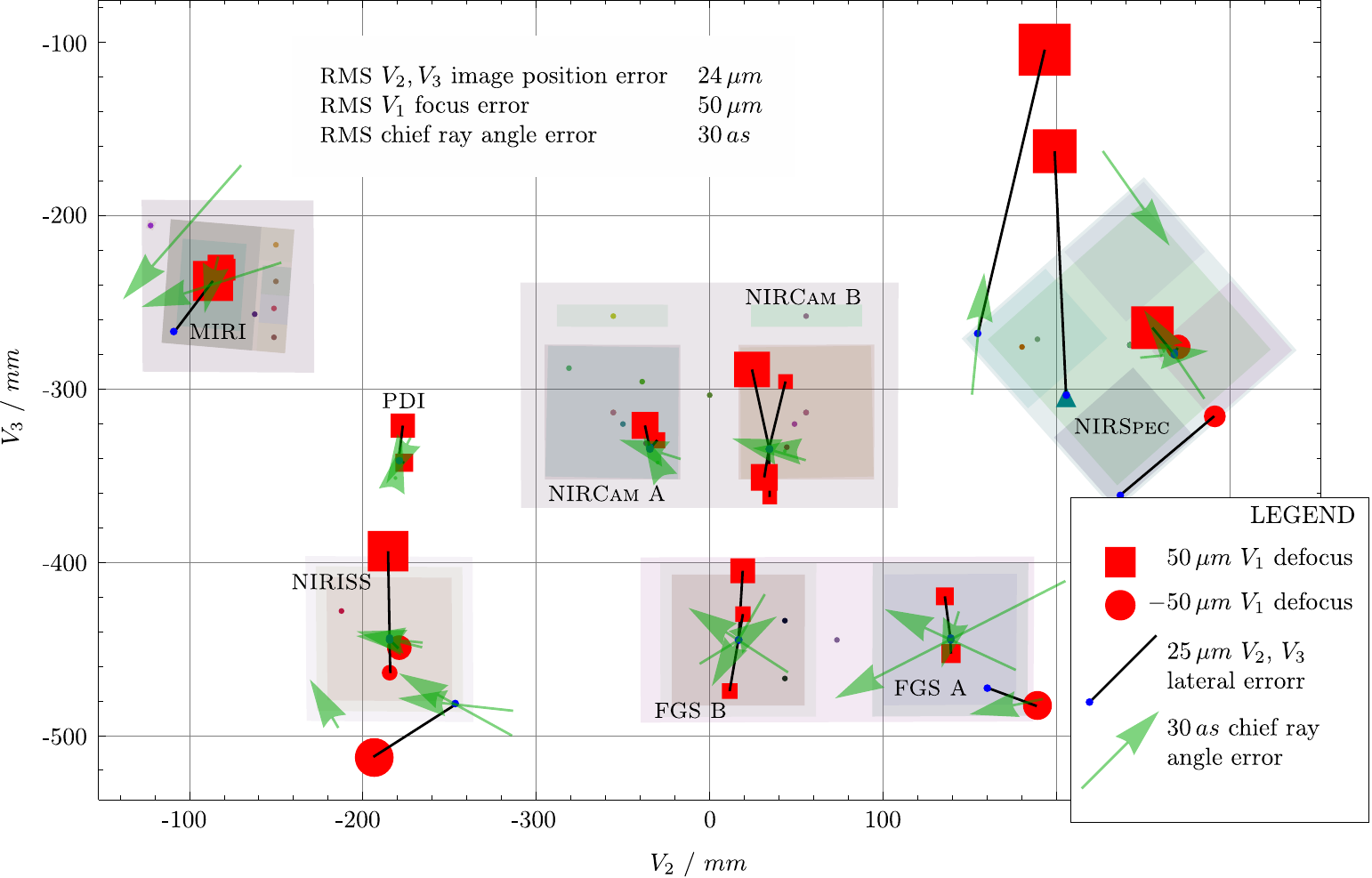}
\caption{\label{fig:BIAResiduals}Sample fitting residuals between measured image positions and chief ray angles and prescription predictions.}
\end{center}
\end{figure}

Figure~\ref{fig:BIAResiduals} shows a sample set of fit residuals for \acr{BIA} data.
Scale factors for each of the three data types (lateral image position, focus and chief ray angle errors) are set independently, so that the errors are clearly visible.
Multiple measurements were accumulated at a number of the field points, so that the measurement noise level can also be seen.
The \acr{RMS} residuals in this case are in good agreement with expected measurement noise and pointing requirements.
As with curve fitting, random scatter of residuals at the level of measurement noise is the most desirable outcome.  
It represents a high-quality fit without systematic errors.
When systematic errors are present, it's an indication that further work is required to refine the test or model until convergence is achieved.

\begin{figure}
\begin{center}
\includegraphics[width=4.2in]{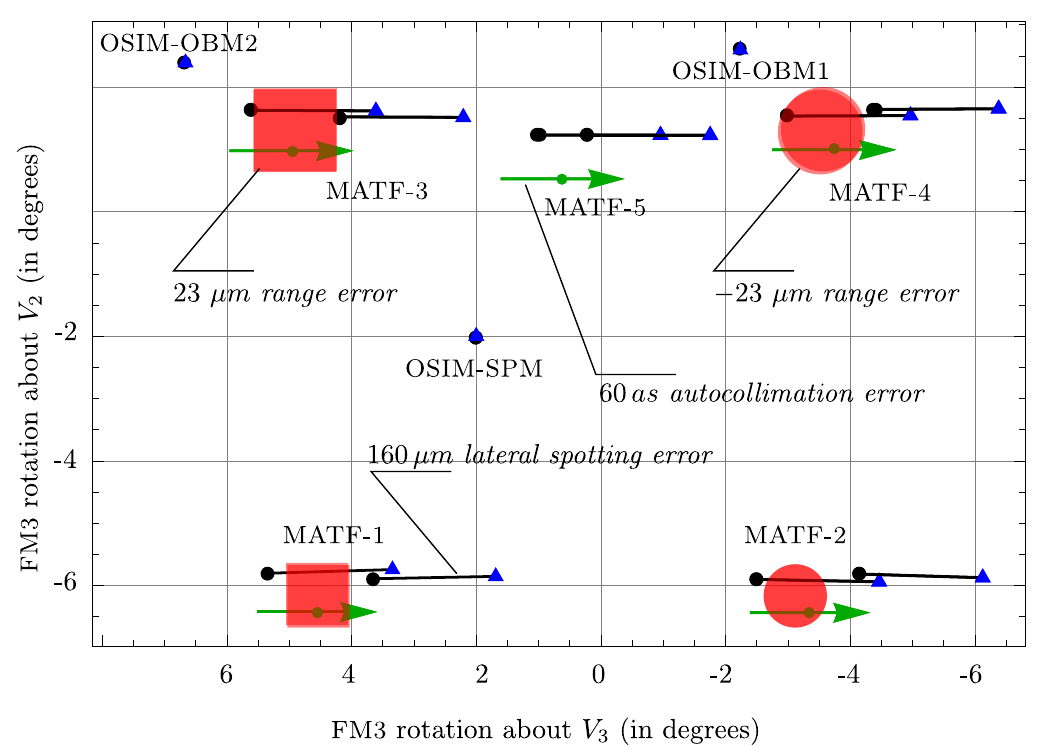}
\caption{\label{fig:ADMResiduals}Synthetic \acr{ADM} residuals for a modeled prescription error in which \acr{OSIM} is rotated $60\,as$ about the $+V_3$ axis. The horizontal axis runs right-to-left in order to match the parity of the \acr{OTE} focal surface.}
\end{center}
\end{figure}

\subsection{\acr{ADM} data and fitting}
In the case of \acr{ADM} measurements, we use a similar scheme.
Here the data types are lateral errors in spotting readings, ranging errors and autocollimation errors.
For spotting errors, we ray trace the measured line-of-sight to a spotting target (such as a backlit pinhole) using our prescription.
We are also given metrologized coordinates for that pinhole, which are derived from photogrammetry and other measurements of the target assemblies.
Then we can calculate a lateral offset of the known target from the traced line-of-sight following Table~\ref{tbl:VectorRayTracing}.
Range errors are similarly computed by comparing the modeled optical path length to a ranging target with the value reported by the 
\acr{ADM}.  Autocollimation errors are obtained by comparing flat mirror normals inferred from alignment telescope autocollimation readings and ray tracing with metrology values.
These errors are dimensionally equivalent to the three error types encountered in \acr{BIA} calibration.
In some cases, they are even driven by the same error modes in the prescription.
Hence we use black lines to represent spotting errors, red squares or circles for ranging errors, and green arrows for autocollimation errors (Figure~\ref{fig:ADMResiduals}).
These are arrayed in \acr{FM3} angle space, with one axis flipped to correspond roughly to the \acr{OTE} focal surface.

This arrangement gives a sense of how the targets would appear if one were to look up through \acr{OSIM} from the \acr{ADM} to the image surface.
Figure~\ref{fig:ADMResiduals} shows a modeled case in which \acr{OSIM} is rotated by $60\,as$ about the $+V_3$ axis.
With external (\acr{MATF}) targets, this rotation couples directly into the autocollimator readings, and also into lateral spotting errors with a scaling factor given by the lever arm between the center of rotation and the target locations.
It also manifests in the range readings as a tilt, with positive errors (red boxes) toward one side and negative errors (circles) on the other.
The internal \acr{OSIM} targets show no effect from the perturbation, as expected.
It's useful to keep a library of \acr{ADM} signatures such as these for every perturbation mode to be applied to a prescription during fitting, for use in deciphering residual plots.
As with most non-linear fitting problems, global optimization is difficult and oversight by an intelligent operator to find a good starting point and guide convergence is a practical necessity.

\subsection{\acr{OSIM} calibration framework}

\begin{figure}
\begin{center}
\includegraphics[width=5.5in]{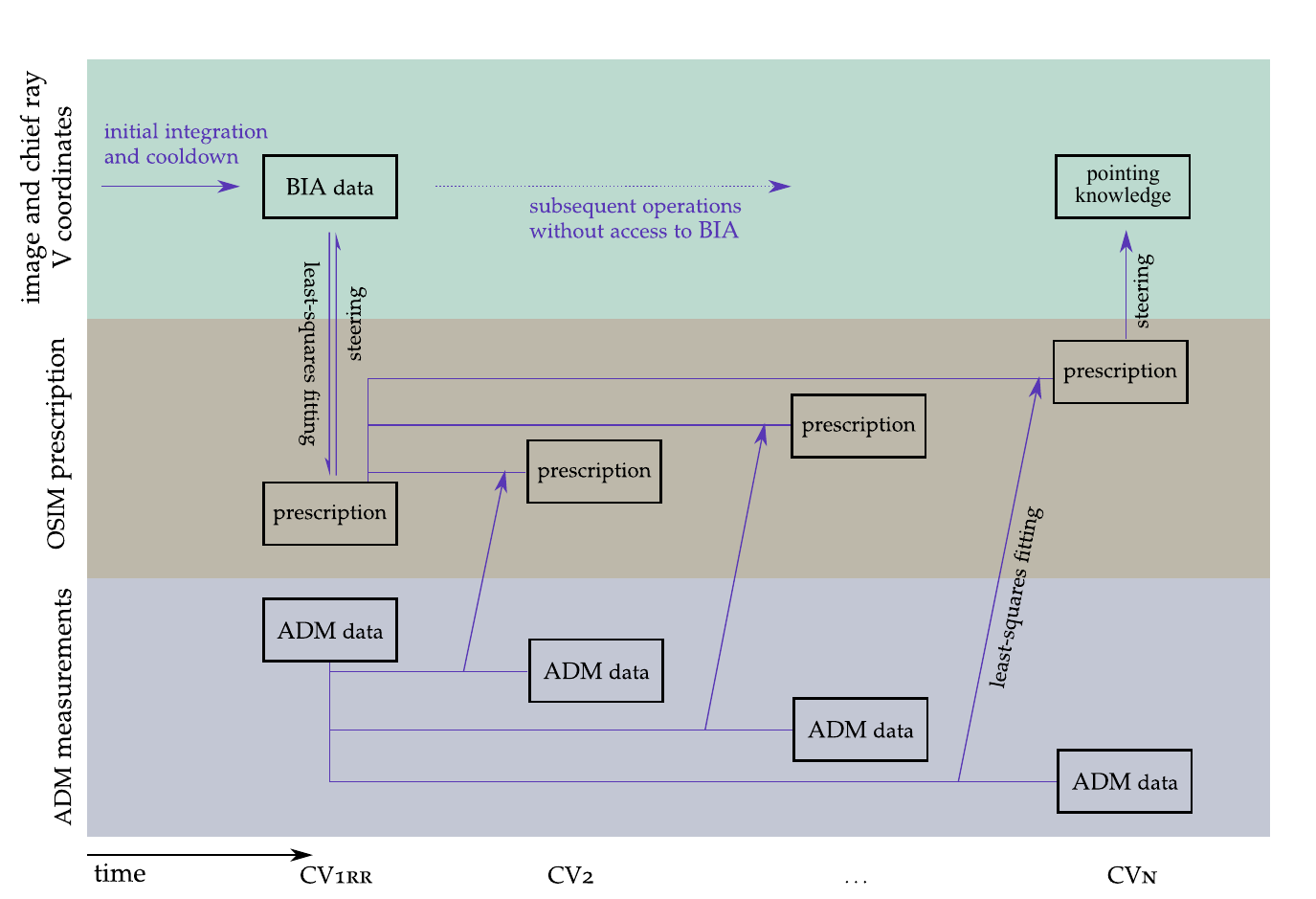}
\caption{\label{fig:PointingSpacesFlow}Maintaining and using \acr{OSIM}'s pointing calibration is an exercise in traversing 3 different data spaces: ray and exit pupil data, prescription data, and \acr{ADM} data.}
\end{center}
\end{figure}

\acr{BIA} and \acr{ADM} measurements are used together to achieve and maintain accurate pointing capability as illustrated in Figure~\ref{fig:PointingSpacesFlow}.
Initial calibration measurements are made with the \acr{BIA} in place (together with baseline \acr{ADM} observations).  
The \acr{BIA} was heavily metrologized at ambient using a number of methods, including laser tracker and laser radar measurements.
Its translation actuation was calibrated under cryo-vacuum operating conditions by means of in-chamber photogrammetry cameras.
Thus absolute coordinate calibration of \acr{OSIM} pointing derives from this reference metrology.
Since neither the \acr{BIA} nor photogrammetry system can be present during \acr{ISIM} testing, the \acr{ADM} is used for transfer measurements, to detect prescription changes which may have been introduced by thermal cycling and loading in the interim since the original calibration.

Within limits, \acr{ADM} data can be used to infer prescription changes which result in differences in readings from the \acr{ADM} target suite and correct the prescription.
The dimensions of the problem are worth noting.
Five dimensions of readings are possible from each complete target assembly --- 2\acr{D} autocollimation errors + 2\acr{D} lateral spotting errors + 1\acr{D} range error.
In a typical measurement campaign, we get data from about 7 target groups, giving a total of about 35 measurements.
However, the \acr{OSIM} prescription is of the order of 400-dimensional, meaning that it takes this many numbers to specify the prescription completely.
These values include not only surface vertex coordinates and normal vectors, but also source coordinates and actuator calibration data (including translation and rotation axes, zero locations and step sizes).
Thus the problem of prescription inference is severely under determined.

The difficulty is partially mitigated by virtue of the fact that some prescription dimensions are degenerate with respect to pointing.
For example, since the \acr{ADM} line-of-sight and \acr{OSIM} ray bundles follow the same path through \acr{FM1}, \acr{FM2} and \acr{FM3} as the image-forming ray bundles, it's not necessary to infer motions of these mirrors independently, but only their net effect.
Suppose a perturbation of \acr{FM1} were incorrectly attributed to \acr{FM2}.
The adjustment made to \acr{FM2} would be such as to have the same optical effect on both the \acr{ADM} and \acr{OSIM} image-forming ray bundles, and so the mis-attribution to \acr{FM2} would be moot as far as pointing is concerned.
Even in cases where the mis-attribution does matter, how much it matters is assessed in terms of pointing outcomes rather than prescription error.
\pagebreak 
By design, the \acr{ADM} lines-of-sight share as much path length as possible with the \acr{OSIM} ray bundles to maximize this degeneracy.
The \acr{ADM}-driven prescription update process has been validated through Monte Carlo analysis and testing (in a blind test in which the \acr{BIA} was present to confirm the final pointing).
So far in the \acr{ISIM} test campaign, prescription updates have not been necessary.  
\acr{OSIM} is proving very stable, which the \acr{ADM} confirms with null results.
            
\section{results from early \acr{ISIM} testing\label{Sect:Results}}
A number of test results are to be had either directly from the steering routines or with application the methods described here.
\subsection{field maps}
\begin{figure}
\begin{center}
\includegraphics[width=6.5in]{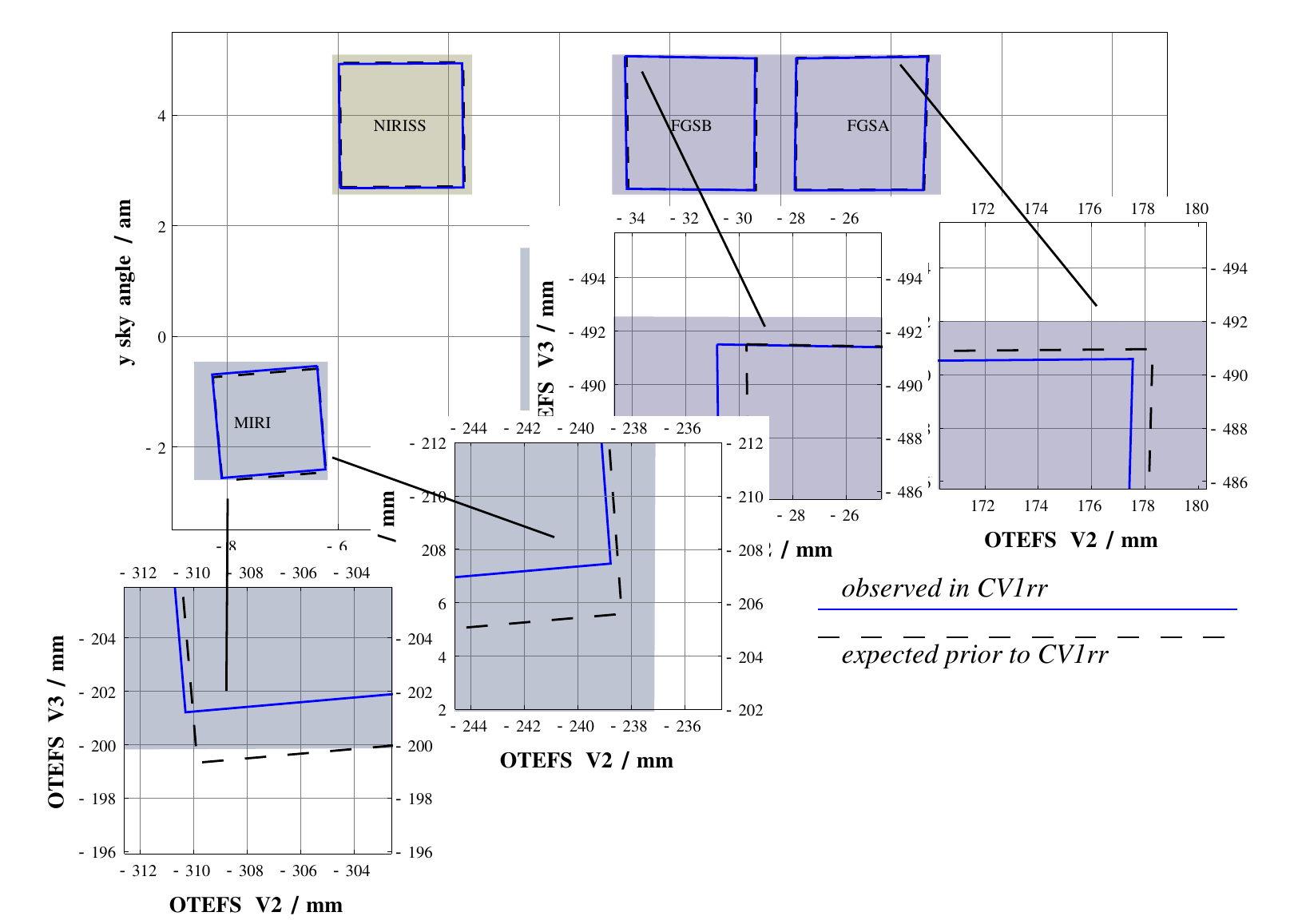}
\caption{\label{fig:CV1RRAllocationCheck}Once calibrated to test data, the field mapping infrastructure of the \acr{OSIM} steering code provides a quick way to estimate science instruments' fields-of-view on the sky for comparison to allocations.}
\end{center}
\end{figure}

During \acr{ISIM} testing, spot images are collected across every science instrument detector array to ensure accurate pointing from \acr{OSIM}.
If necessary, transforms in the network shown in Figure~\ref{fig:CoordMapNetwork} are updated by least-squares fitting.
This allows subsequent tests to be carried out with pixel-scale pointing accuracy, easily locating image plane features (such as slits and microshutters) and working around bad pixels in engineering-model detectors.
The field mapping infrastructure provides a ready method to extrapolate these test points to locate the detector corners, and map them out to sky coordinates for comparison to allocations.
Figure~\ref{fig:CV1RRAllocationCheck} shows such results for the science instruments present in the \acr{CV1rr} cryo-vac test.

\subsection{pupil imaging module}
\begin{figure}
\begin{center}
\includegraphics[width=4.39in]{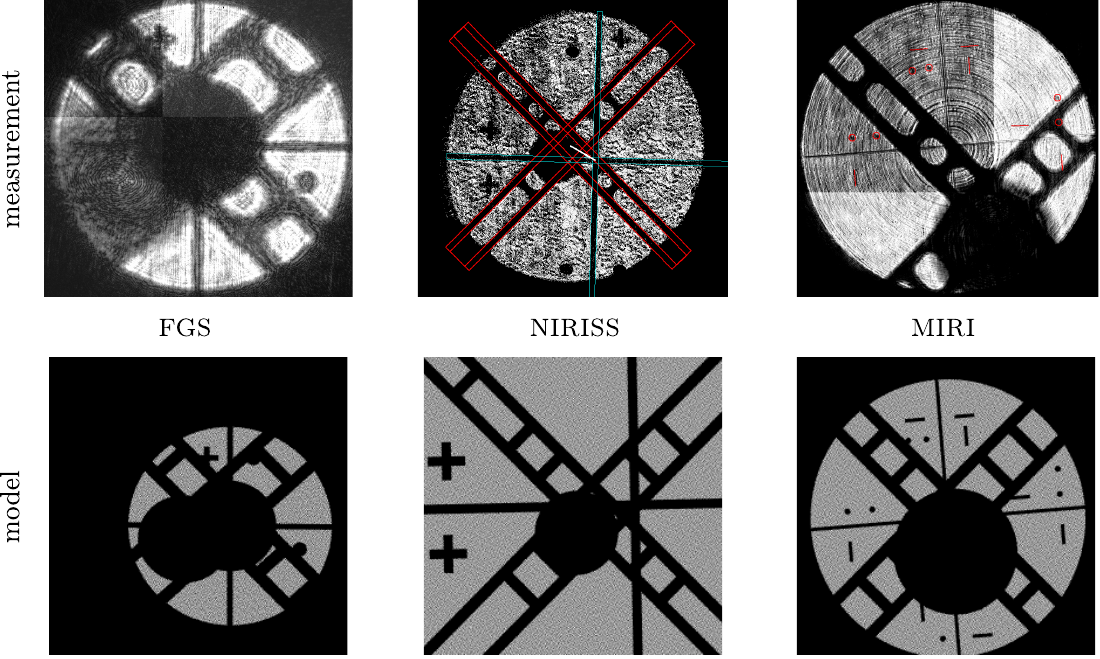}
\caption{\label{fig:PIMResults}Pupil imaging module images (top row) showing overlapped fiducials from the \acr{OSIM} and science instrument pupil alignment reticles in \acr{CV1rr}. In the \acr{NIRISS} case, overlaid feature masks used in image analysis are also shown. The bottom row shows simulated images (in these cases using geometric optics calculations only) obtained with application of the steering code to \acr{Code V} optical models.}
\end{center}
\end{figure}

Figure~\ref{fig:PIMResults} shows real \acr{PIM} images for the science instruments present in \acr{CV1rr}, together with modeled versions.
Considerable modeling and breadboarding was carried out during \acr{OSIM} development to study the impact of diffraction ringing (which is clearly visible in the top row of the figure) and investigate image analysis techniques.
The simulated images shown are straight forward geometric versions, meant for quick comparison in order to advise analysts with regard to image orientation, obscurations from the source injection apertures in \acr{FM1}, and \acr{PAR} fiducials.
They were generated with the aid of steering utilities which map as-run encoder settings from the image headers into \acr{Code V} models for this sort of analysis.  
The same models could be used for diffraction-based analyses as well.

The analysis workflow involves an operator who works in a graphical image processing program to align template shapes to the \acr{PAR} fiducials in the images. 
The final coordinates of the templates are then analyzed to give quantitative offsets between pupils.  
Bookkeeping of a number of expected offsets is required.
These include as-built coordinates of the \acr{PAR}s, field-dependence of the \acr{OTE} exit pupil location, and corrections for expected ground-to-orbit differences.
Note that the modeled images in Figure~\ref{fig:PIMResults} do not necessarily incorporate all such effects, so differences along columns in the figure are not necessarily pupil errors.  
It is clear from the modeled images however that perfect overlap of \acr{PAR} crosshairs is \emph{not} expected.

\subsection{point diffraction interferometer}
The \acr{PDI} provides a wealth of information.
Beyond \foreign{in situ} measurement of wavefront error, it provides the ability to directly image the pupil, estimate chief ray angle, and provide a common point of reference for image position.
Pupil images may be be acquired without interference fringes by slightly offsetting the input beam (using \acr{FM3} or the \acr{SPM}) from the pinhole.
Pupil illumination is a significant consideration in phase retrieval, so this capability is important for wavefront error knowledge.

\begin{figure}
\begin{center}
\includegraphics[width=5.64in]{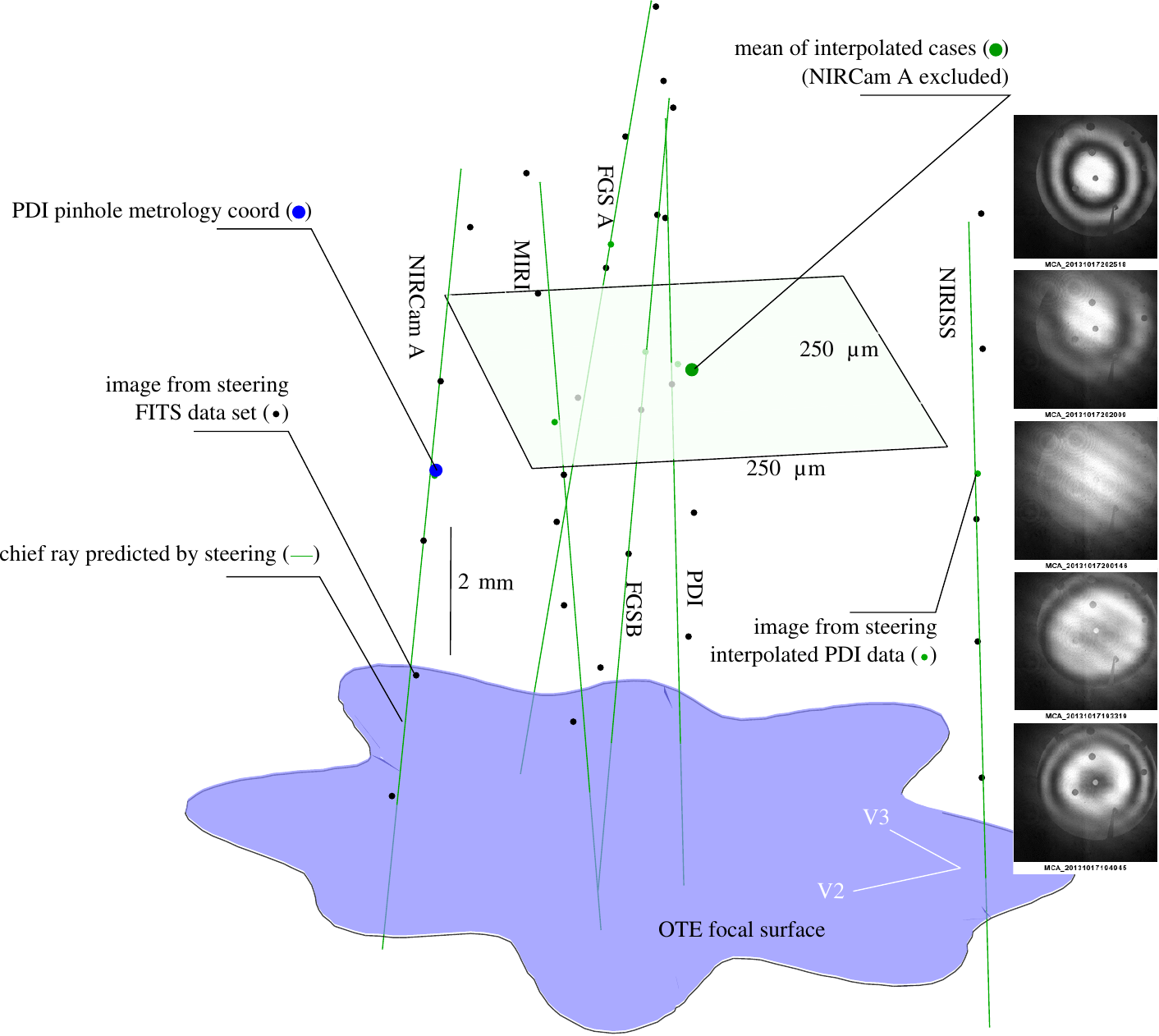}
\caption{\label{fig:PDISteeringCrosscheck}Image positions predicted by the \acr{OSIM} steering code run on through-focus sweeps at the \acr{PDI} with a selection of sources at disparate field locations.  The rays would ideally converge at the pinhole location, which is the observed result to within relevant tolerances.  (Note that the lateral dimensions are scaled differently than the axial dimension in order to make residual errors evident.)}
\end{center}
\end{figure}

Since a lateral offset of the \acr{OSIM} image from the \acr{PDI} pinhole produces tilt fringes, the \acr{PDI} provides a stable image location for steering cross checks (Figure~\ref{fig:PDISteeringCrosscheck}).
Each source was steered in turn to the \acr{PDI} and manually adjusted to produce a null fringe.
In practice this meant performing a through-focus sweep and interpolating the analyzed wavefronts to obtain an accurate null.
The steering code and current \acr{OSIM} prescription was then used to predict the image coordinates for the as-run encoder settings.
\enlargethispage{-\baselineskip}
These steered coordinates were compared for internal consistency and to the metrologized pinhole coordinate, yielding agreement at the expected level.
 
The fact that the chief rays (shown in green) in Figure~\ref{fig:PDISteeringCrosscheck} match the sequence of image points for each sweep is a further check on the prescription.  It indicates that the \acr{SPM}'s translation axes are accurately known.  This is important in practice to enable through-focus phase retrieval sweeps without beam wander, particularly for instruments with narrow field restrictions such as the fixed imaging windows and shutter apertures in \acr{NIRSpec}.

\section{\foreign{ad hoc} and serendipitous cross-checks}
The presence of the \acr{ADM} affords a number of additional cross-checks, some of which were discovered serendipitously during integration and test.  
For instance, one gives us the ability to bridge the separation between the beam paths of the \acr{ADM} and \acr{OSIM}'s output images.

\subsection{direct viewing of the \acr{OSIM} spot}
Since light obeys a principle of superposition, there is no fundamental constraint which locks the \acr{ADM} line-of-sight to the \acr{OSIM} master chief ray (\acr{MCR}).
The \acr{ADM}'s alignment telescope comes very close to solving this problem.  
Using the \acr{ADM}'s internal \acr{RM} mirrors to align the line-of-sight through two points simultaneously is sufficient to boresight the \acr{ADM} to the ray through the points.
In fact, we have two alignment targets on the master chief ray (\acr{MCR}) --- the \acr{MCR} optical fiber on the source plate and a target in the pupil wheel.
In this case there is a remaining complexity, however, in that  the \acr{ADM}'s beam path does not reflect from the primary mirror.
By aligning the \acr{ADM} to the pupil as seen downstream of the primary mirror and the source as seen directly, we would in effect be assuming that the \acr{MCR} meets the primary mirror at normal incidence.  

To test that assumption, we found during test that we could substitute a different known point on the \acr{OSIM} ray.  
The idea was to place a screen of some sort at the \acr{OSIM} image, and then sight the spot with the \acr{ADM}.
In combination with sighting to the pupil, this gives two points on the chief ray, and in this case both are on the downstream side of the primary mirror, eliminating the need for an assumption of normal incidence.

In practice, we weren't able to see the spot from a diffuse screen, presumably because insufficient light was directed back to the \acr{ADM}.
However, we had good results with other types of return elements, such as ground corner cube vertices and even the phase retrieval detector on the \acr{BIA}.
Small dithers of image position were used to confirm that the spot was in good focus and not being returned by specular reflection.
It's an interesting question as to why specular reflection (from the corner cube or a normal mirror) does not work for this exercise.  
They would in fact return the beam back into \acr{OSIM}, but the obscuration from the \acr{ADM}'s aperture in the center of the primary mirror would be mapped back to the \acr{ADM}, so no light would be received.
(Using a tilted flat mirror works, but ensuring that the beam is focused on the surface of the mirror to avoid displacement of the image is difficult.)

\begin{figure}
\begin{center}
\includegraphics[width=5.22in]{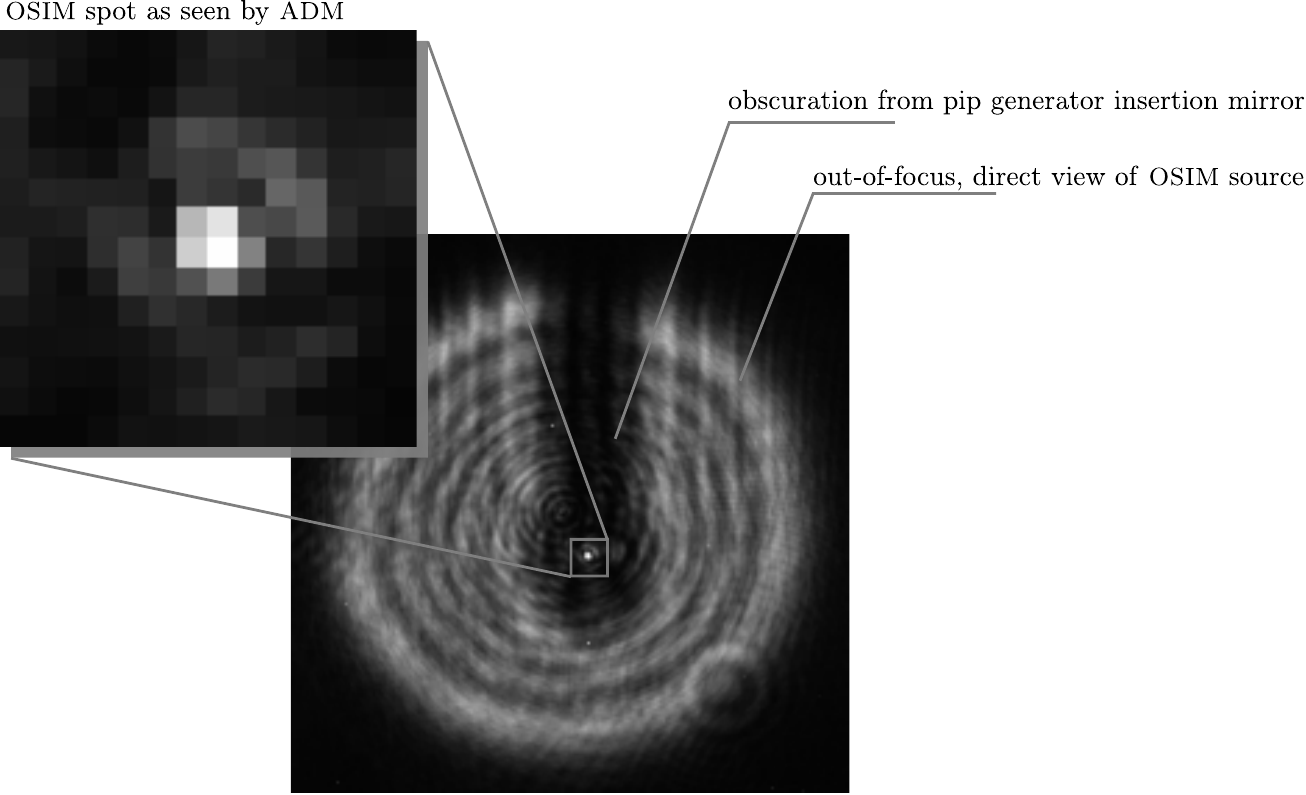}
\caption{\label{fig:SpotOSIMSpot}To cross-check pointing calibration or boresight the \acr{ADM} to the \acr{OSIM} chief ray, \acr{OSIM}'s master chief ray beam can be focused onto a  target which glints back into the optical system, and the spot sighted directly with the alignment telescope. Direct illumination from the \acr{OSIM} source would be problematic, except that the alignment telescope has a pip generator whose insertion mirror obscures the out-of-focus spot.}
\end{center}
\end{figure}

There is another subtle complexity of this test.
When viewing the \acr{OSIM} image, the object (\ie the fiber source on the source plate) is also directly in view, albeit out of focus.
The considerable stray light received from this direct path would likely swamp the image, except that it lies conveniently in the shadow of an insertion mirror for a pip generator installed on the alignment telescope (Figure~\ref{fig:SpotOSIMSpot}).

\subsection{boresighting the alignment telescope to the phase retrieval camera}
\begin{figure}
\begin{center}
\includegraphics[width=5.94in]{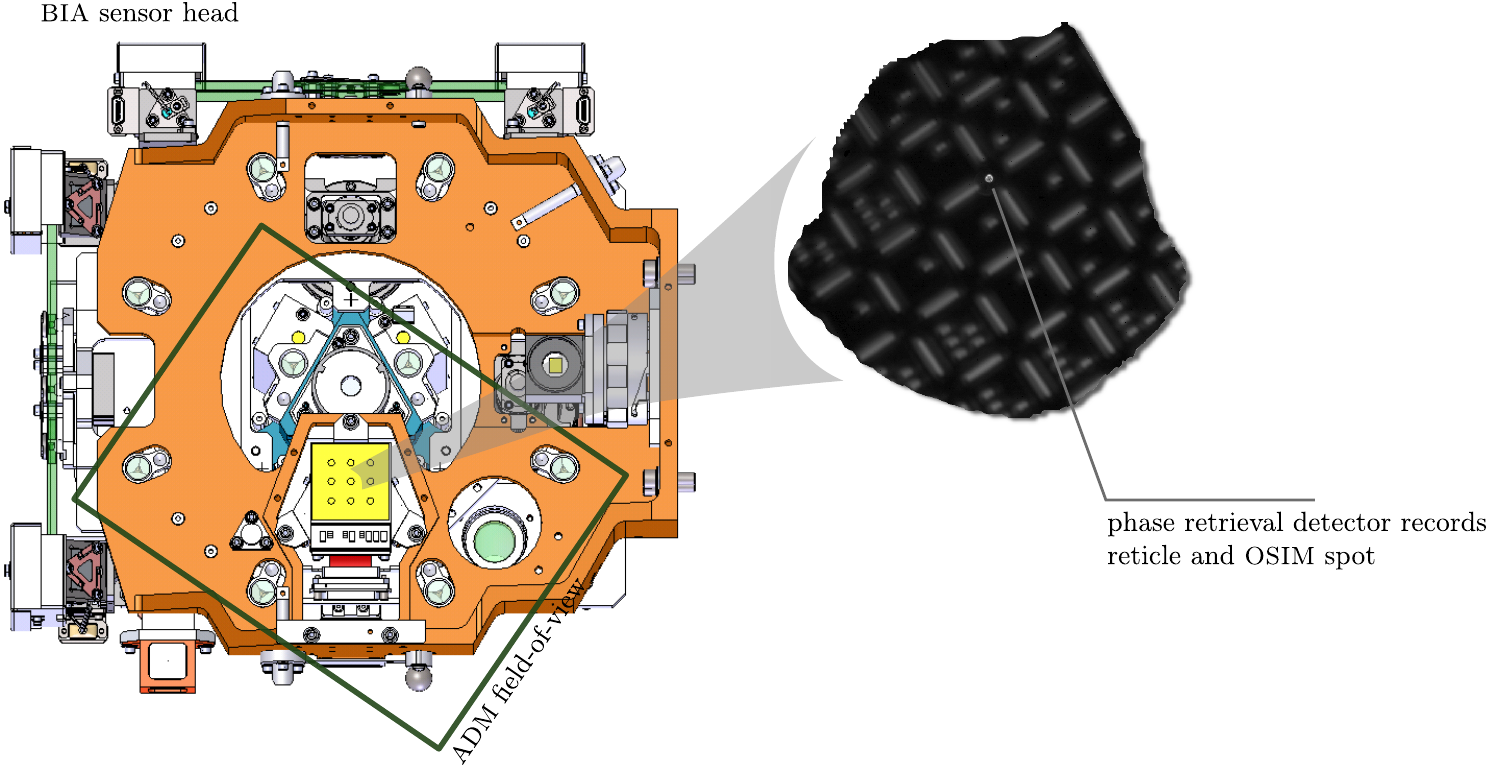}
\caption{\label{fig:BIALevitonReticle}During testing with the \acr{BIA}, the phase retrieval camera captured the projected image of the \acr{ADM}'s reticle, allowing a cross-check of the sensor head and translation encoder calibration against prescription knowledge for \acr{OSIM} and the \acr{ADM}. Firing \acr{OSIM}'s master chief ray source also allows a check of the \acr{ADM} boresight to \acr{OSIM}.}
\end{center}
\end{figure}

The reticle in the \acr{ADM}'s alignment telescope autocollimator is projected downstream, providing other opportunities for cross checks.
For instance, when testing with the \acr{BIA} we were able to focus the alignment telescope to the \acr{OTE} focal surface and position the \acr{BIA}'s phase retrieval detector to record the reticle image (Figure~\ref{fig:BIALevitonReticle}). 
The \acr{BIA}'s translation actuators and sensor head features are metrologized to provide knowledge of the \acr{V} coordinates of pixels in the phase retrieval camera.  
Locating the reticle boresight in the image recorded by the phase retrieval sensor then provides a cross-check of sensor head metrology and prescription knowledge for \acr{OSIM} and the \acr{ADM}. 

\subsection{boresighting the point diffraction interferometer}
With the \acr{ADM} alignment telescope focused to the \acr{OSIM} pupil, the \acr{PDI} can similarly image the autocollimator reticle.
This is useful for determining or confirming the boresight pixel of the \acr{PDI} on the \acr{BIA} \foreign{in situ}. 
In this scenario, the \acr{PDI} boresight is defined by the normal vector of its entrance reference flat (or where a ray through the pinhole parallel to that vector traces to on the detector).
Boresight is defined in this way since the reference flat is accessible to the \acr{ADM}'s autocollimator.
By exercising \acr{FM3}, the \acr{ADM} can autocollimate from the flat on the \acr{PDI} and metrologized flats on the \acr{MATF} in turn, and use the recorded \acr{FM3} motion to  refer the \acr{PDI} reference mirror normal to \acr{V} coordinates.
Capturing an image of the reticle directly with the \acr{PDI} then provides a mean to refer pixel coordinates in pupil images to chief ray angles in \acr{V} coordinates.

\section{\acr{FRED} stray light modeling}
\begin{figure}
\begin{center}
\includegraphics[width=3.04in]{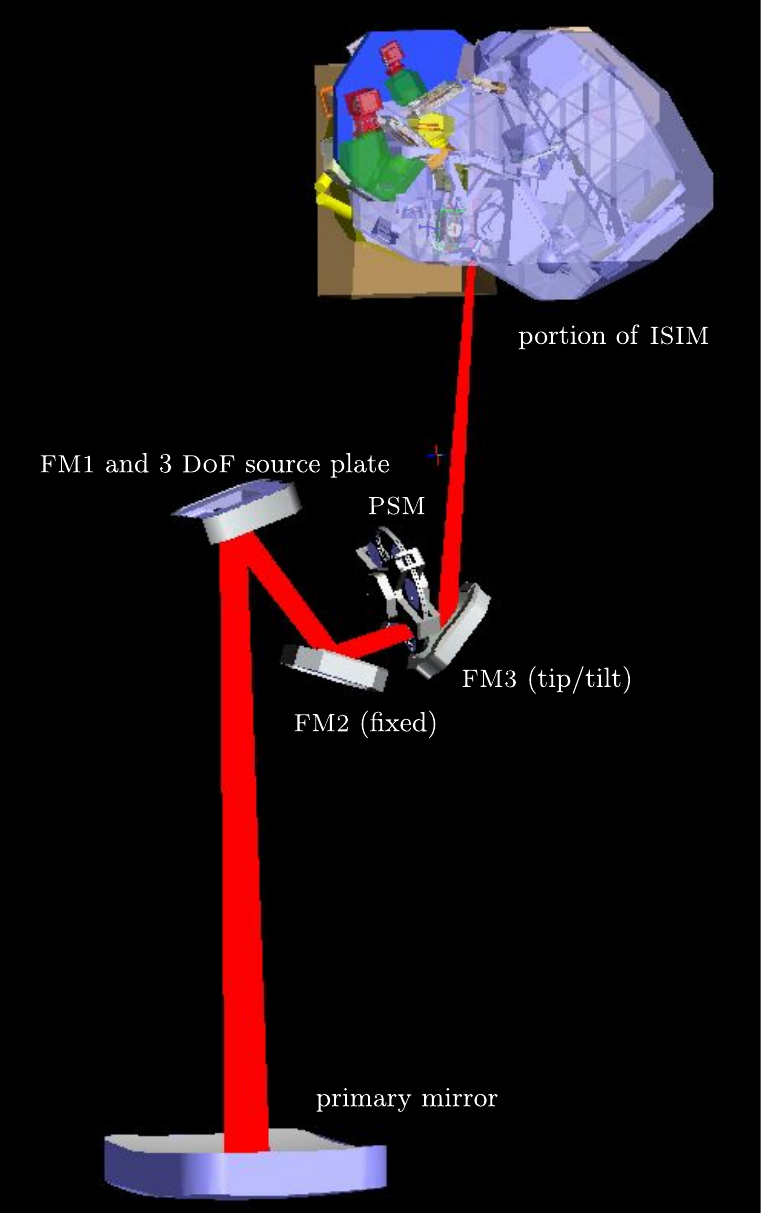}
\caption{\label{fig:StrayLightHardware}Relevant \acr{OSIM} and \acr{ISIM} hardware in the stray light model.  Much of the model is hidden in this view for clarity.}
\end{center}
\end{figure}

A significant --- though initially unanticipated --- use of the pointing algorithm is in the \acr{ISIM} stray light model. In order to accurately predict stray light artifacts in the \acr{CV} optical testing performed at the \acr{ISIM} level, and to easily differentiate between artifacts that would be visible only during testing and not on-orbit, a single stray light model containing \acr{ISIM}, \acr{OSIM}, and the supporting hardware in the \acr{ISIM} \acr{CV} tests was built in the \acr{FRED} Optical Engineering\texttrademark software package. A screen capture showing just the relevant optical elements of \acr{OSIM} is shown in Figure~\ref{fig:StrayLightHardware}, with the \acr{OSIM} elements underneath and \acr{ISIM} above. 

The original intent for the stray light model was to integrate the \acr{OSIM} steering code directly into the \acr{FRED} scripting language so that the end user did not need to rely on any external sources to operate the model. The complexity of the steering code led to the eventual decision to call the python code executables from the \acr{FRED} scripts. This allowed for continued steering code development that did not need to be replicated within \acr{FRED}, which would have effectively forked the code bases. To make the use of the steering code more efficient for the \acr{FRED} scripts, an extension was developed which output a comma separated value text file instead of a \acr{Code V} sequence file. 

The resulting combination of \acr{FRED} scripts reading the \acr{CSV} output from the \acr{OSIM} steering code yields a highly efficient means of updating the stray light model prior to performing a ray trace. The previous method of generating a \acr{Code V} model and then manually translating optical element positions within the \acr{FRED} model took on the order of an hour per \acr{OSIM} configuration. The scripted method allows the user to update the model in a matter of seconds, and also allows for batch lists of \acr{OSIM} configurations to be generated. This changed the paradigm of stray light modeling from targeting a very few general cases to allowing users to investigate any number of field points and \acr{ISIM} configurations, limited only by \acr{CPU} time. 

\begin{figure}
\begin{center}
\includegraphics[width=6.1in]{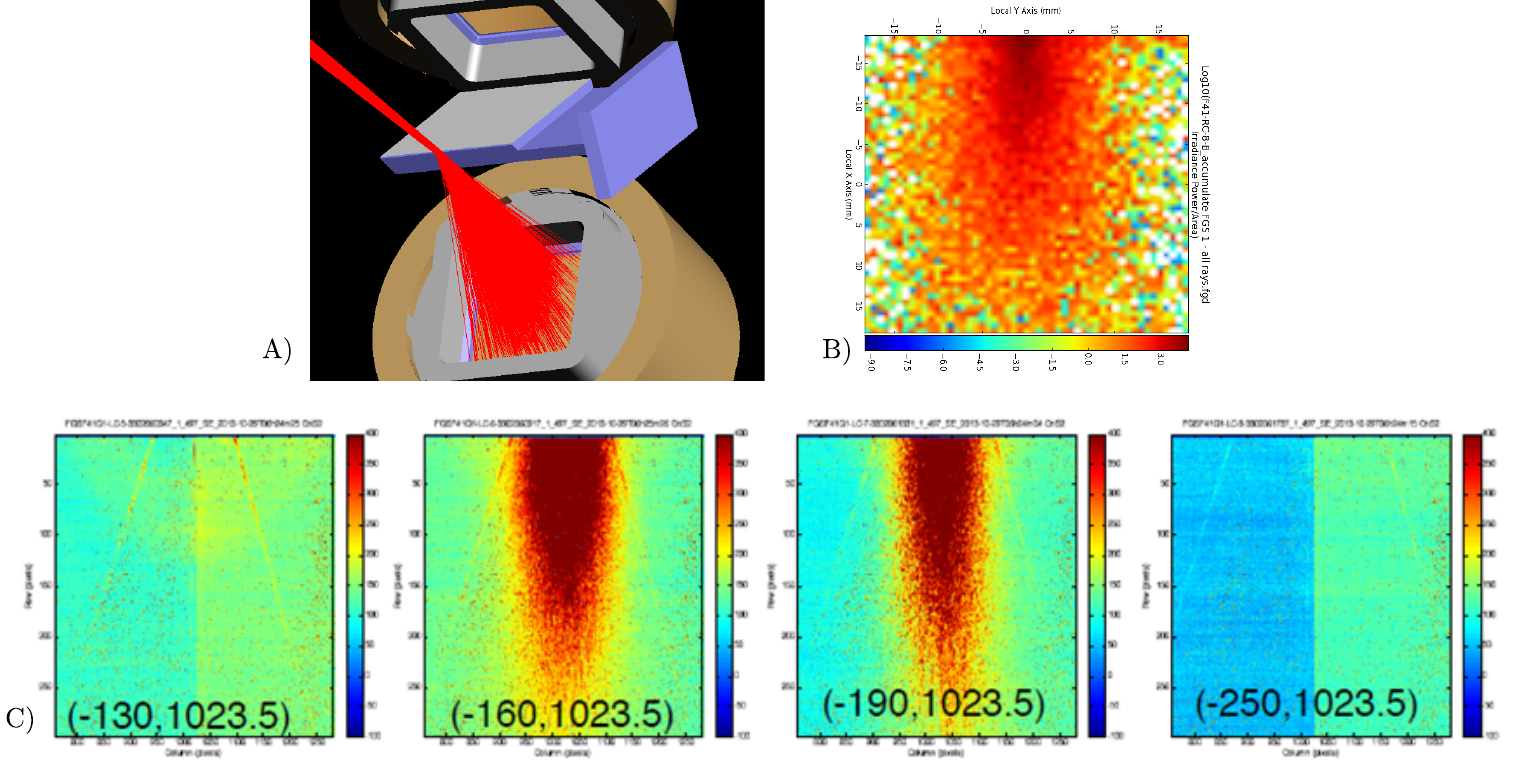}
\caption{\label{fig:FGSPickOffScatter} Ray paths between the \acr{FGS1} and \acr{FGS2} fields are expected to scatter from the \acr{FGS} Detector Fold Mirror onto the \acr{FGS1} detector (A).  Observed images (such as the example in B) agree well with simulated images (such as the sweep sequence in C). }
\end{center}
\end{figure}

\begin{figure}

\begin{center}
\includegraphics[width=4.82in]{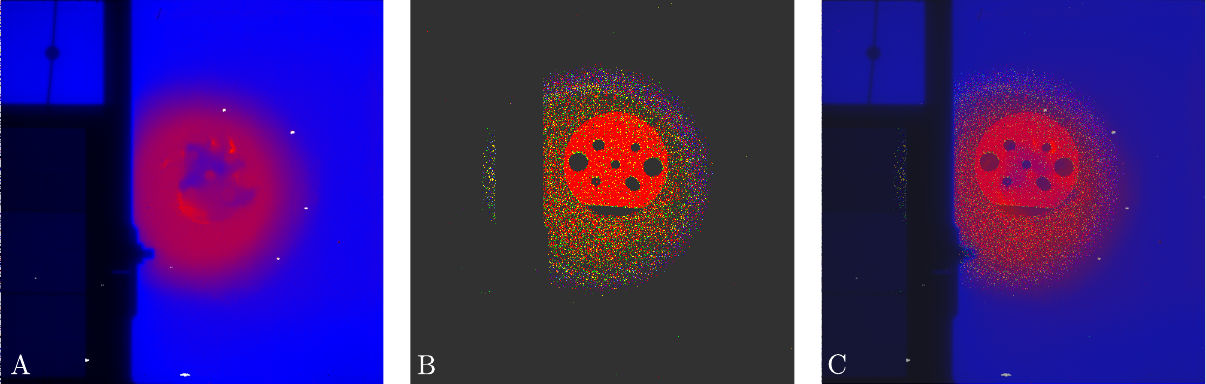}
\caption{\label{fig:SFBackground}A comparison of thermal background from the 100~K superferrule and source plate in a representative \acr{MIRI} dark image, including an observed image (A), the modeled  self-emission (B) and an overlay of the two (C). }
\end{center}
\end{figure}

The \acr{OSIM} steering code allowed users to predict a few very important stray light artifacts prior to the \acr{ISIM} \acr{CV1rr} test, and perform near-live-time diagnosis of artifacts discovered during the test. A few examples of the model correlation to artifacts observed in the \acr{ISIM} \acr{CV1rr} test campaign follow.
Figure~\ref{fig:FGSPickOffScatter} illustrates a study of stray light expected due to scatter from the \acr{FGS} detector pick-off mirror for field points between the allocations for the two channels.
Figure~\ref{fig:SFBackground} shows thermal emission from the \acr{OSIM} source superferrules as recorded by \acr{MIRI}.  Although the background temperature of \acr{OSIM} is merely 100~K, this represents a significant thermal emission source for \acr{MIRI} and was one of the first images available during initial cooldown with \acr{ISIM}.  
Modeling confirmed that the geometry of the imaged superferrule and out-of-focus illumination from the baffle enclosing its aperture in \acr{FM1} were observed as expected.
The variation in the emission from the face of the superferrule is also expected, as the ``cap'' surface was machined and then welded to the cylinder, yielding variations in emissivity.

\begin{figure}
\begin{center}
\includegraphics[width=4.67in]{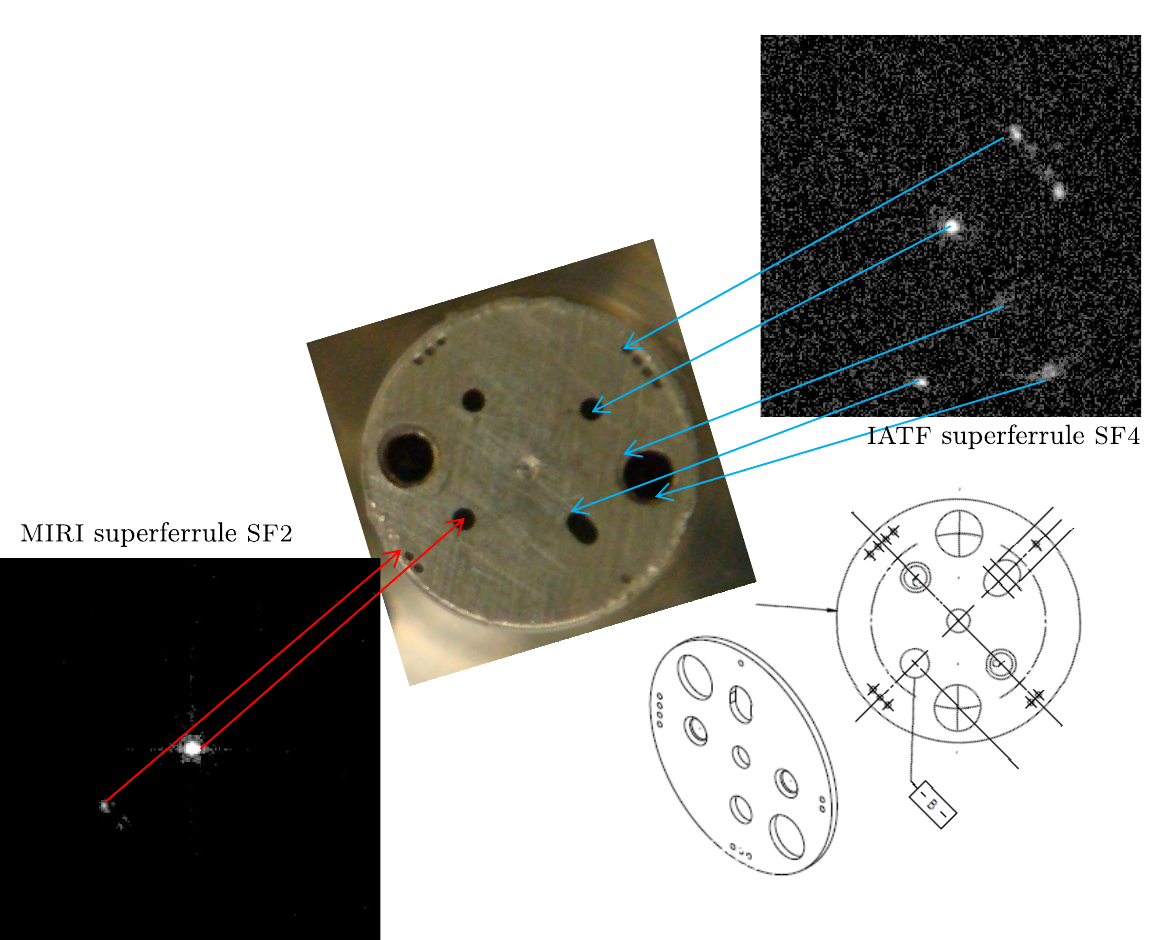}
\caption{\label{fig:MIRISuperferrule} Stray light observed in \acr{MIRI} images taken with the \acr{MIRI} subfiber 2 and \acr{IATF}subfiber 4. Artifacts were traced to fiducial holes in the surface of the superferrule that allowed light to leak out of the superferrule.}
\end{center}
\end{figure}

\begin{figure}
\begin{center}
\includegraphics[width=4.93in]{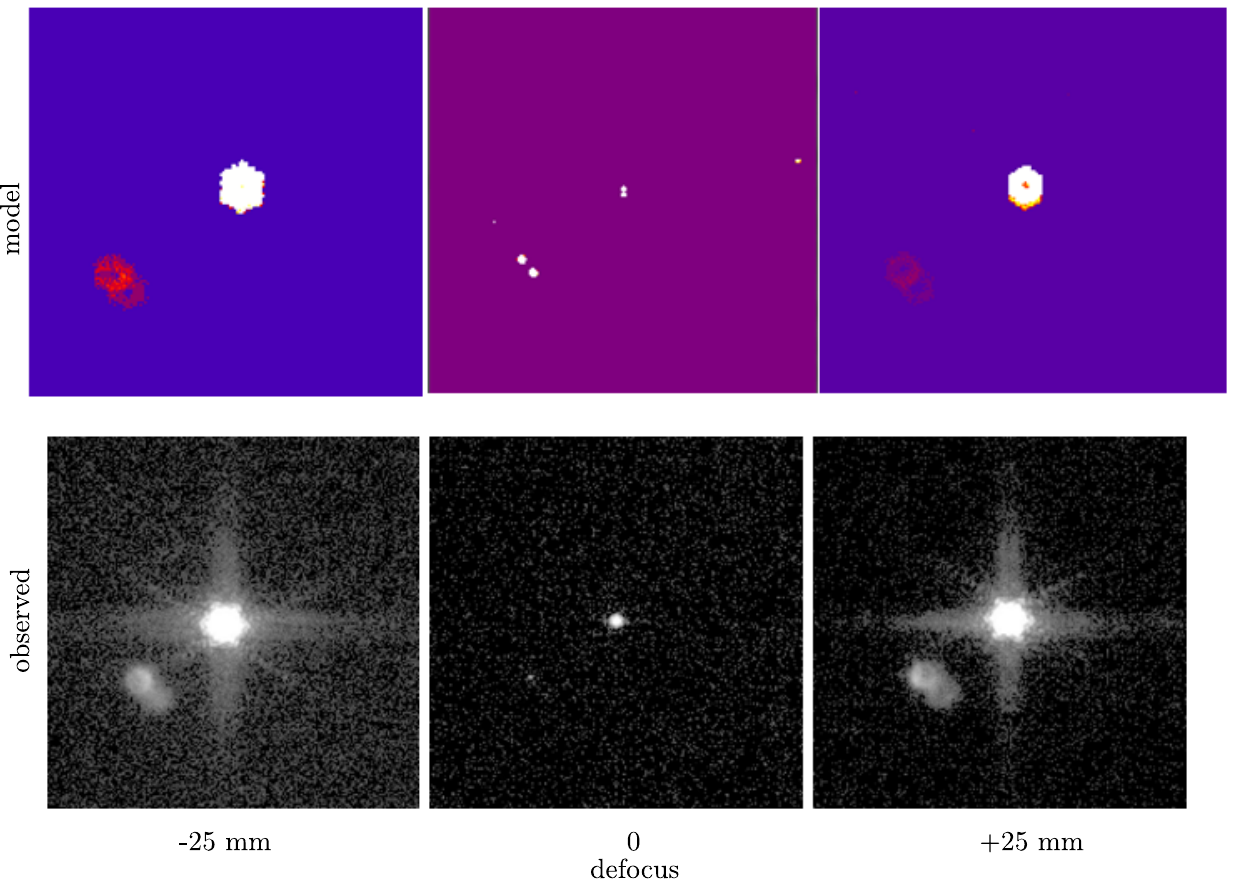}
\caption{\label{fig:MIRIGhosts} The stray light model of \acr{ISIM} and a steerable \acr{OSIM} enabled easy simulation of focus sweeps. Excellent correlation was obtained between the modeled (lower row) and observed (above row) ghost artifacts.}
\end{center}
\end{figure}

Figure~\ref{fig:MIRISuperferrule} shows modeling which was carried out to identify spurious illumination observed with certain \acr{MIRI} sources.  The face plate of the superferrule had been marked with small fiducial through holes. Light leakage was visible with recessed sources.  Once identified, this stray light did not impact test results. 
Figure~\ref{fig:MIRIGhosts} compares modeled and observed results of a study of ghosts in the \acr{MIRI} optical design.


\clearpage

\section{conclusions}
By way of a summary, we list the following best practices which our experience with ray tracing in \acr{OSIM} leads us to advocate. 
\begin{quote}
\begin{description}
 \item[Communicate with vectors] (\ie direction cosines) rather than Euler angles when specifying direction vectors and coordinate systems
 \item[Make units, coordinate system and (for position vectors) origin explicit] in variable names.
 \item[Use a \texttt{GenSys} routine] rather rather than cross-products to construct coordinate systems based on known vectors
 \item[Plot residual errors on their own scale] to make systematic effects visible.
 \item[Stand on the shoulders of giants] by capitalizing on existing tools. Excellent algorithms (pseudo-inversion, shortest path in a graph) and software packages (python and associated packages, Gimp, ImageMagick, Inkscape, Subversion) are available, many under open licenses with cross-platform support which enable wide and rapid development and dissemination of a project's tools. 
\end{description}
\end{quote}

\section*{acknowledgements}        
With a program the scale of \acr{JWST}, significant contributors indubitably number in the thousands, making an adequate set of acknowledgements impossible.  
Even so, we would be remiss without mentioning Bill Eichhorn, Jenny Chu and Severine Tournois who made the \acr{PDI} work.  Ray Ohl, Joe Hayden, Theo Hadjimichael and Phil Coulter were instrumental in target metrology for \acr{OSIM} and \acr{ISIM}.
Scott Antonille was the untiring advocate responsible for many of the innovations described here.
Rob von Handorf, Jeff Kirk, Brad Greeley, Clint Davis and Doug Leviton made \acr{OSIM} work, Randal Telfer was key in calibrating it, and Ed Shade, Corbett Smith, Pam Davila, Robert Rashford and Paul Volmer got it built. 


\bibliography{JrnAbbrev,GeneralBibliography}   

\begin{thebibliography}{10}
\newcommand{\enquote}[1]{``#1''}

\bibitem{Lightsey2012}
P.~A. Lightsey, C.~Atkinson, M.~Clampin, and L.~D. Feinberg, \enquote{{James
  {W}ebb {S}pace {T}elescope: large deployable cryogenic telescope in space},}
  Optical Engineering \textbf{51}, 011,003--1--011,003--19 (2012).

\bibitem{Davila2004}
P.~S. Davila, B.~J. Bos, J.~Contreras, C.~Evans, M.~A. Greenhouse, G.~Hobbs,
  W.~Holota, L.~W. Huff, J.~B. Hutchings, T.~H. Jamieson, P.~A. Lightsey,
  C.~Morbey, R.~Murowinski, M.~J. Rieke, N.~Rowlands, B.~Steakley, M.~Wells,
  M.~B. te~Plate, and G.~S. Wright, \enquote{{The James Webb Space Telescope
  science instrument suite: an overview of optical designs},} Proc. SPIE
  \textbf{5487}, 611--627 (2004).

\bibitem{Kimble2012}
R.~A. Kimble, P.~S. Davila, C.~E. Diaz, L.~D. Feinberg, S.~D. Glazer, G.~S.
  Jones, J.~M. Marsh, G.~W. Matthews, D.~B. McGuffey, P.~H. O'Rear, D.~D.
  Ramey, C.~A. Reis, S.~C. Texter, and T.~L. Whitman, \enquote{The integration
  and test program of the {J}ames {W}ebb {S}pace {T}elescope,} Proc. SPIE
  \textbf{8442}, 84,422K (2012).

\bibitem{Davila2008}
{Pamela S. Davila \textit{et al.}}, \enquote{{The Optical Telescope Element
  Simulator for the {J}ames {W}ebb {S}pace {T}elescope},} Proc. SPIE
  \textbf{7010}, E1--12 (2008).

\bibitem{Sullivan2010}
J.~Sullivan, B.~Eichhorn, R.~von Handorf, D.~Sabatke, N.~Barr, R.~Nyquist,
  B.~Pederson, R.~Bennnett, P.~Volmer, D.~Happs, A.~Nagle, R.~Ortiz, T.~Kouri,
  P.~Hauser, J.~Seerveld, D.~Kubalak, B.~Greeley, C.~Hakun, D.~Leviton,
  Q.~Gong, P.~Davila, R.~Ohl, J.~Kirk, C.~Davis, J.~Chu, E.~Wilson, B.~Chang,
  S.~Mann, R.~Rashford, and C.~Smith, \enquote{Manufacturing and integration
  status of the {JWST} {OSIM} optical simulator,} Proc. SPIE \textbf{7731},
  77,313V (2010).

\bibitem{Sabatke2009}
D.~{Sabatke}, R.~{von Handorf}, and J.~{Sullivan}, \enquote{{Polarization and
  fold mirrors in application of the Leica Absolute Distance Meter},} Proc.
  SPIE \textbf{7461}, 74610N (2009).

\bibitem{Leviton2003}
D.~B. {Leviton}, J.~{Kirk}, and L.~{Lobsinger}, \enquote{{Ultrahigh-resolution
  Cartesian absolute optical encoder},} Proc. SPIE \textbf{5190}, 111--121
  (2003).

\bibitem{Nowak2010}
M.~D. Nowak, P.~E. Cleveland, E.~Cofie, J.~A. Crane, P.~S. Davila, B.~H.
  Eegholm, R.~P. Hammond, J.~B. Heaney, J.~E. Hylan, J.~D. Johnston, R.~G. Ohl,
  J.~D. Orndorff, D.~L. Osgood, K.~W. Redman, H.~P. Sampler, S.~A. Smee, J.~M.
  Stock, F.~T. Threat, R.~A. Woodruff, and P.~J. Young, \enquote{Cryogenic
  performance of a high precision photogrammetry system for verification of the
  {J}ames {W}ebb {S}pace {T}elescope {I}ntegrated {S}cience {I}nstrument
  {M}odule and associated ground support equipment structural alignment
  requirements,} Proc. SPIE \textbf{7793}, 77,930A--9 (2010).

\bibitem{Hadjimichael2010}
T.~Hadjimichael, D.~Kubalak, A.~Slotwinski, P.~Davila, B.~Eegholm, W.~Eichhorn,
  J.~Hayden, E.~Mentzell, R.~Ohl, G.~Scharfstein, and R.~Telfer,
  \enquote{Cryogenic metrology for the {J}ames {W}ebb {S}pace {T}elescope
  {I}ntegrated {S}cience {I}nstrument {M}odule alignment target fixtures using
  laser radar through a chamber window,} Proc. SPIE \textbf{7793}, 77,930B--10
  (2010).

\bibitem{BarrettImageScience}
H.~H. Barrett and K.~J. Myers, \emph{{Foundations of Image Science}} (Wiley,
  Hoboken, New Jersey, 2004).

\bibitem{Dorst2007}
{Leo Dorst}, {Daniel Fontijne}, and {Stephen Mann}, \emph{{Geometric {A}lgebra
  for {C}omputer {S}cience}} (Morgan Kaufmann, San Francisco, 2007).

\bibitem{Python}
\emph{{Python programming language official website}},
  \href{http://www.python.org/}{http://www.python.org/}.

\bibitem{CohenTannoudji1977}
C.~Cohen-Tannoudji, B.~Diu, and F.~Laloe, \emph{{Quantum Mechanics}}, vol.~1
  (Wiley, 1977).

\bibitem{Mallat1999}
S.~Mallat, \emph{{A Wavelet Tour of Signal Processing}}, {Electronics \&
  Electrical} (Academic Press, 1999).

\bibitem{Golub1996}
G.~Golub and C.~{Van Loan}, \emph{{Matrix Computations}}, {Johns Hopkins
  Studies in the Mathematical Sciences} (Johns Hopkins University Press, 1996).

\bibitem{Smith2007}
W.~Smith, \emph{{Modern Optical Engineering, 4th Ed.}}, {McGraw Hill
  professional} (McGraw-Hill Education, 2007).

\bibitem{Numpy}
\emph{{Numpy package official website}},
  \href{http://www.numpy.org/}{http://www.numpy.org/}.

\bibitem{Scipy}
\emph{{Scipy package official website}},
  \href{http://www.scipy.org/}{http://www.scipy.org/}.

\bibitem{NetworkX}
\emph{{NetworkX package official website}},
  \href{http://networkx.lanl.gov/}{http://networkx.lanl.gov/}.

\end{thebibliography}
\bibliographystyle{osajnl}   

\end{document}